\documentclass[11pt]{article}
\usepackage{graphicx}
\usepackage{amssymb}
\usepackage{epstopdf}
\usepackage[]{amsmath,amssymb}
\usepackage{amsfonts}
\usepackage{graphics,epsfig}
\def\ie{{\rm i.e.,\/}\ }
\def\etc{{\rm etc.\/}\ }

\def \otimesdot {\stackrel{\cdot}{\otimes}}
\def\one{\mbox{\rm 1}\hskip-2.8pt \mbox{\rm l}}

\newcommand{\ZZ}{\mathbb{Z}}

\begin{document}

\title{On quantum symmetries of the non--ADE graph ${F_4}$}
\author{R. Coquereaux, E. Isasi\footnote{Supported, in part, by a fellowship of "Fundaci\'on Gran Mariscal
de Ayacucho", Venezuela} \\
           {\it Centre de Physique Th\'eorique - CNRS} \\
             {\it Campus de Luminy - Case 907}           \\
             {\it F-13288 Marseille - France}      }

\date{}
\maketitle
\baselineskip=11.6pt

\begin{abstract}
We describe quantum symmetries  associated with the $F_4$ 
Dynkin diagram. Our study stems from an analysis of the (Ocneanu) modular
splitting equation applied to a partition function which is invariant under
a particular congruence subgroup of the modular group.  
\end{abstract}

\begin{flushleft}{CPT-2004/P.045}\\
\end{flushleft}

\section{Introduction} 

Modular invariant partition functions of the affine $SU(2)$ conformal 
field theory models have been classified long ago \cite{CapItzZub};  they  
follow an $ADE$ classification. 
There are three infinite series called $A_n$, $D_{2n}$, $D_{2n+1}$ 
and three exceptional cases
called $E_6$, $E_7$ and $E_8$. The terminology was justified from the 
fact that exponents of the corresponding Lie groups appear in the 
expression of the corresponding partition functions but, 
originally, this labelling using Dynkin diagrams was only a name 
since the diagram itself was not an  ingredient in the construction. 
Later, and following in particular the work of \cite{Pasquier}\cite{Pasquier:alg}, the 
corresponding field theory models were built directly from the data 
associated with the diagrams themselves. 

About ten years ago, the occurrence of  $ADE$ diagrams in the affine 
$SU(2)$ classification was understood in a rather different way.
One observation (already present in reference \cite{Pasquier}) is 
that the vector space spanned by
the vertices of a diagram $A_n$ possesses an associative and 
commutative algebra structure encoded by the diagram itself : this 
so-called ``graph algebra'' has a unit $\tau_0$ (the first vertex), 
one generator $\tau_1$ (the next vertex) and multiplication of a 
vertex $\tau_p$ by $\tau_1$ is given by the sum of the neighbors of 
$\tau_p$. So, $\tau_1 \tau_p =  \tau_{p-1}  + \tau_{p+1}$ when $p < 
n-1$ and $\tau_1 \tau_{n-1}  = \tau_{n-2}$. The structure constants 
of this algebra (which can be understood as a quantum version of the  
algebra of $SU(2)$ irreps at roots of unity) are positive integers.
  In some cases ($A_n$, $D_{2n}$, $E_6$, $E_8$), the vector space spanned by the vertices of a chosen 
Dynkin diagram of type $ADE$
  also enjoys self-fusion, \ie admits, like $A_n$ itself,  an 
associative algebra structure, with  structure 
constants that are positive integers, together with a multiplication table  ``related'' to the 
graph algebra of the corresponding $A_n$.
Another important observation is that this vector space 
 is always a module over the 
graph algebra of $A_n$ where $n+1$ is the Coxeter number of the 
chosen diagram. For instance the vector spaces spanned by vertices 
of  the diagrams $E_6$ and $D_7$ are modules over the graph algebra of 
$A_{11}$ (their common Coxeter number is $12$).
 The Ocneanu construction \cite{Ocneanu:paths} associates with every $ADE$ Dynkin 
diagram $G$ a special kind of  weak Hopf algebra (or quantum 
groupoid). This bialgebra $B(G)$ is finite dimensional and 
semi-simple for its two associative structures. 
 Existence of a coproduct on the underlying vector space (and on its 
dual) allows one to take tensor product of irreducible 
representations (or co-representations) and decompose them into 
irreducible
 components. One obtains in this way two -- usually distinct -- 
algebras of characters. The first,  called the fusion algebra of $G$, 
and denoted $A(G)$,  can be identified with the graph algebra of 
$A_n$, where  $n+1$ is the Coxeter number of $G$ (this number is 
defined in a purely combinatorial way, and does not require
any reference to the theory of Lie algebras or Coxeter groups).  The second algebra of 
characters is called the algebra of quantum symmetries and denoted 
$Oc(G)$; it is an associative -- but not necessarily commutative --  
algebra with two  (usually distinct) generators. It comes with a 
particular basis, and the multiplication of its basis elements by the 
two generators is encoded by a graph called the Ocneanu graph of 
$G$.  The quantum groupoid associated with $G$ is ``special''  in 
several ways.  For instance,  each one of the two algebras $A(G)$ and 
$Oc(G)$ is a bimodule over the other; this bimodule structure is 
encoded by a set of ``toric matrices'' and one of them can 
be identified with the modular invariant partition function for a 
physical system. The others, which are not modular invariant, can be 
physically interpreted, in the framework of Boundary Conformal Field 
Theory, as partition functions  in the presence of defects, see 
\cite{PetZub:Oc}.  Physical applications of this general formalism may
be found in statistical mechanics \cite{Mercat-et-al},  string theory or quantum gravity, but this
is  not the subject of the present paper.
 
 Modular invariance can be investigated either directly, by 
associating vertices $\tau_p$ of the graph $A_n$ with explicit 
functions $\chi_p$ defined on the upper - half plane (the characters 
of an affine Kac-Moody algebra), or more simply, by checking
commutation relations between a particular toric matrix and the
 two generators $S$ and $T$ of the $SL(2,\ZZ)$ group in a particular 
representation (Verlinde - Hurwitz, \cite{Hurwitz} \cite{Verlinde} ). The matrix $S$, as given by the 
Verlinde formula,  is actually a non-commutative analog of the table 
of characters for a finite group and can be obtained directly from 
the multiplication table of  the graph algebra of $A_n$.
 
 The theory that was briefly summarized  above can be 
generalized from $SU(2)$ to $SU(N)$ or,  more generally, to any 
affine algebra associated with a chosen Lie group. The familiar 
simply laced $ADE$ Dynkin diagrams are associated with the 
affine $SU(2)$ theory, but more general Coxeter-Dynkin systems (each 
system being a collection of diagrams together with their 
corresponding quantum groupoids) can be obtained \cite{Ocneanu:MSRI} . For instance the 
 Di Francesco - Zuber diagrams  \cite{oldZuber} \cite{DiFZub}  are associated with the 
$SU(3)$ system.
  
 What notion comes first ? The  chosen diagram, member of some 
Coxeter-Dynkin system ? Its corresponding quantum groupoid ? Or the 
associated modular invariant ? The starting point may be
a matter of taste... However, from a practical point of view, it is 
probably better to  start from the combinatorial data provided by 
a given modular invariant. Indeed, apart from $SU(2)$ and, to some 
extent,  $SU(3)$, the Coxeter-Dynkin system itself  is not  a priori 
known, whereas existence of several algorithms, mostly due to T. 
Gannon, allows one to explore up to rather high levels the possibly 
new modular invariants associated with every choice of an affine Kac 
- Moody algebra.
The primary data is then a given modular invariant --  a 
sesquilinear form in the characters of some  affine Kac-Moody 
algebra. From the resolution, over the positive integers, of a 
particular equation, called ``equation of modular splitting'' (more 
about it later), one can determine first the set of  toric matrices, 
and then  the algebra of quantum symmetries. The associated
Dynkin diagram -- an $ADE$ diagram in the case of the $SU(2)$ system -- or, more 
generally,  the particular member of some higher Coxeter-Dynkin system becomes an 
{\sl outcome}Ê of the construction, not a starting point. The $SU(3)$ system of diagrams 
was already essentially determined by \cite{oldZuber}, \cite{DiFZub}, and recovered 
by A. Ocneanu, using this modular splitting technique, together with 
the known classification of $SU(3)$ modular invariants obtained by 
T.Gannon \cite{Gannon}.  It was also 
 the route followed by \cite{Ocneanu:Bariloche}  in his  
classification of the $SU(4)$ system. 
In all these examples, starting from a modular invariant partition function, 
one obtains a diagram that  
is simply laced (\ie an $ADE$ diagram, for the  $SU(2)$ system), or is a
generalization of what could be defined as a ``simply laced diagram'', for
the higher systems.

In the present paper we are only interested in the affine $SU(2)$ 
system and we  want to start from a partition function which is not 
modular invariant but which is nevertheless invariant under some 
particular congruence subgroup. 
Our starting point is the so-called  ``$F_4$ partition function'' which
appears as a kind of $\ZZ_2$ orbifold of the  $E_6$ modular invariant
(its name comes from the fact that  exponents of the Lie group
$F_4$ appear in its expression).  After the work of \cite{Dubrovin},  this partition function was discussed in \cite{Zuber:1993vm}.
 Our purpose is neither to propose any physical conformal field theory model that would lead to this expression
nor to investigate its analytical properties
but rather to analyze the  equation of  modular splitting 
corresponding to this  particular choice of a modular non-invariant expression 
and see what algebraic structure it gives.
 Starting from this data, we shall  determine, in turn, a set of toric matrices and 
an algebra of quantum symmetries (described by an Ocneanu graph). Actually we
do not even suppose that we ``know'' what the diagram $F_4$ is :
 it will appear as a subgraph of the graph of quantum symmetries. 
 We shall not try here to generalize our study to cover other non-ADE 
cases of the $SU(2)$ system, and shall not investigate, either, 
the non simply laced diagrams belonging to higher systems. Since we want to 
focus  on the modular splitting technique itself, we shall not try to use or modify, in this
non-simply laced case, the definitions of the product and coproduct  that usually lead to a quantum groupo\" \i d structure, but  the fact that one can  solve the equation of modular splitting, define 
toric matrices and determine an algebra of 
quantum symmetries together with two algebras of characters obeying the usual
quadratic sum rules  is by itself  a non trivial result that suggests
further developments.
 
 The algorithms that we use in order to solve the relevant 
equations and determine the algebraic quantities of interest have been developed by
the authors, but it may well be that 
more efficient  techniques have been found by others\footnote{The results presented long ago in \cite{Ocneanu:paths} or \cite{Ocneanu:Bariloche}, for instance, require the use of analogous techniques. }.  If so, this 
information is not available.  Although devoted to the study of a particular member of an 
unusual class (a non-$ADE$ example), we believe that the present 
paper may be of interest for the reader who wants to see
how the general technique based on the equation of modular splitting works, since it does 
not seem to be  documented in the  literature.  Apart from 
considerations of computing efficiency, it should not be too hard to adapt the 
following analysis  to study other cases, simply laced or not, associated with 
any Coxeter-Dynkin system\footnote{See the forthcoming paper \cite{Schieber:SU4} where an exceptional simply laced example of the $SU(4)$ system is studied.}.

Before ending this introduction, let us stress the fact that the 
present paper does not require any knowledge of the theory of Lie 
algebras or quantum groups (or quantum groupoids). All the algebras
used in the following are associative algebras, not Lie. A sentence 
like ``the algebra $A_n$'' means actually ``the associative graph 
algebra corresponding to the diagram $A_n$'' and we usually identify 
a given Dynkin diagram with the vector space or the $\ZZ$-module spanned by its
vertices.

\section{Toric matrices from modular splitting} 

\subsection*{Invariance of the partition function}

For the $SU(2)$ system, there are three modular invariant partition 
functions at Coxeter number $\kappa = 12$:  they are respectively associated 
with the diagrams $A_{11}$, $D_{7}$ and $E_6$. The first  case (also 
called ``diagonal'') is given by
$$
Z_{A_{11}} =\sum_{n=0}^{10}\left|\chi_n\right|^2
$$ 
Modular invariance of this expression can be explicitly checked by 
performing the transformations $T : \tau \rightarrow \tau +1$ and $S: 
\tau \rightarrow -1/\tau$; to do that, one can use an explicit 
expressions for the eleven characters $\chi_p$ of affine $SU(2)$ at 
level $10$, 
for instance in terms of theta functions. Another possibility, which is 
simpler, is to represent $S$ and $T$ by $11 \times 11$ matrices, 
namely $S_{ij} = \sqrt{\frac{2}{\pi}} sin(\pi \frac{(i+1)(j+1)}{\kappa})$,
$0\leq i, j \leq \kappa - 2$, 
and $T_{ij}= exp[2 i \pi (\frac{(j+1)^2}{4 \kappa} - \frac{1}{8})] \delta_{ij}$, and check that they commute with the 
modular matrix $M$ associated with $Z$ (the relation between the two 
being of course, $Z = \overline{\chi} M \chi$).  Commutation is 
obvious since $M$ is the diagonal unit matrix $\one_{11}$.

The $E_6$ partition function is given by
$$
Z_{E_6} 
=\left|\chi_0+\chi_6\right|^2+\left|\chi_3+\chi_7\right|^2+\left|\chi_4+\chi_{10}\right|^2
$$ 
Its modular invariance can be checked as above but  the 
associated modular matrix is non-trivial.  $Z_{E_6}$ is a sum of 
three modulus squared ``generalized characters''
$\lambda_{1}=\chi_0+\chi_6$, $\lambda_{2}=\chi_3+\chi_7$, and $\lambda_{3}=\chi_4+\chi_{10}.$
$M$  has $12$ non - zero entries equal to $1$. This is related  to 
the fact that the Ocneanu graph has 12 points, three of them being 
ambichiral, the corresponding algebra being commutative.

We now turn to $F_4$, which is our object of study. The partition 
function is
$$
Z_{F_4} =  
\left|\chi_0+\chi_6\right|^2+\left|\chi_4+\chi_{10}\right|^2
$$
The fact that it is {\sl not \/} modular invariant is obvious from 
the modular transformations of the generalized characters of $E_6$:
Under $\tau \rightarrow -1/\tau$, $\lambda_{1} \rightarrow 
(1/2)(\lambda_{1}+\lambda_{2})-(1/\sqrt{2})\lambda_{2}$,   $\lambda_{2} \rightarrow 
(1/\sqrt{2})(\lambda_{3} - \lambda_{1})$,  $\lambda_{3} \rightarrow 
(1/2)(\lambda_{1}+\lambda_{2})+(1/\sqrt{2})\lambda_{2}$.and under $\tau \rightarrow 
\tau +1$, $\lambda_{1} \rightarrow e^{\frac{19 i \pi}{24}} \lambda_{1}$,
$\lambda_{2} \rightarrow  e^{\frac{10 i \pi}{24}} \lambda_{2} $, $\lambda_{3} \rightarrow  
e^{\frac{- 5 i \pi}{24}} \lambda_{3} $.
 However, these relations also show that it is invariant 
under the transformations $\tau\longrightarrow\tau +2$ and 
$\tau\longrightarrow\frac{\tau}{2\tau +1}$ that span a
congruence subgroup\footnote{We do not use the corresponding projective group}  $\Gamma^{(2)}_0$ of $SL(2,\ZZ)$ at level $2$. It is actually easier 
to show this by checking explicitly that the  modular matrix 
associated with $Z_{F_4}$, namely 
{\footnotesize
$$
M=\left(\begin{array}{ccccccccccc}
   1 & . & . & . & . & . & 1 & . & . & . & . \\
   . & . & . & . & . & . & . & . & . & . & . \\
   . & . & . & . & . & . & . & . & . & . & . \\
   . & . & . & . & . & . & . & . & . & . & . \\
   . & . & . & . & 1 & . & . & . & . & . & 1 \\
   . & . & . & . & . & . & . & . & . & . & . \\
   1 & . & . & . & . & . & 1 & . & . & . & . \\
   . & . & . & . & . & . & . & . & . & . & . \\
   . & . & . & . & . & . & . & . & . & . & . \\
   . & . & . & . & . & . & . & . & . & . & . \\
   . & . & . & . & 1 & . & . & . & . & . & 1 \\ \end{array}\right)
$$      
}
commutes with the generators $T^2$ and $ST^{-2}S$ of $\Gamma^{(2)}_0$.
Notice that the exponents of the Lie group $F_4$ appear on the diagonal of $M$, 
hence the name chosen for this
partition function.  From the fact that $Z_{F_4}$ is a sum of two squares, 
one expects two ambichiral points in the Ocneanu graph (indeed it will 
turn out to be so). $M$  has $8$ non - zero entries equal to $1$, 
this could suggest that the corresponding Ocneanu graph has
the same number of vertices... however, as we shall see, this is {\sl 
not \/} so.

\subsection*{Toric matrices (definition)}

Let $m,n,\ldots$ (rather than $\tau_m, \tau_n$, or $\chi_m, \chi_n$) 
denote the vertices of $A_{11}$ and $x,y, \ldots$ the vertices of the 
Ocneanu graph $Oc$ to be found. As already mentioned, $Oc$ should 
be a bi-module over $A_{11}$ so that there exist a collection of 
$11\times 11$ toric  matrices $W_{x,y}$, with positive integer 
entries, such that
$$
m \, x \, n = \sum_y  {(W_{x,y})}_{m,n} \; y
$$
These are ``toric matrices with two twists'' ($x$ and $y$). If $o$ is 
the number of vertices in $Oc$ -- it will be determined later --  the 
number of such matrices is of course $o^2$ but many of them may 
coincide. 
Since $A_{11}$ and $Oc$ play a dual role, it is  useful to introduce 
the $11^2$ matrices $V_{m,n}$ of size $o \times o$  by 
$$
(V_{m,n})_{x,y} = {(W_{x,y})}_{m,n}
$$
The algebra of quantum symmetries has a unit, called $\underline 0$, a 
particular vertex of $Oc$, so that we have also `` matrices with 
one twist'' $W_{0,y}$ and $W_{x,0}$. When the example under
 study corresponds to a simply laced situation (for instance the $ADE$ 
cases of the usual $SU(2)$ system) and  if the algebra $Oc$ is commutative.
one shows that $W_{x,y} = 
W_{y,x}$. However, we are now in 
a new situation and should keep our mind open.  The resolution of the 
general modular splitting  equation will, in any case, determine 
all these  quantities.

\subsection*{Equations of modular splitting}

The general equation of modular splitting expresses associativity of 
a bi-module structure and reads
$$
(m \, n) \, x (\, p \, q) = (m \, (n \, x \, p) \, q) 
$$
The products $(m \, n)$ or  $(\, p \, q)$ belong to the fusion 
algebra, \ie for instance,  to $A_{11}$, and involve its structure constants defined 
by $m \, n = \sum_p N_{m,n}^p \; p$. They are completely symmetric in their three indices.
These positive integers are determined, recursively, by the -- 
truncated -- $SU(2)$ algebra of compositions of spins. For $A_{11}$ 
one obtains $11$ matrices $N_m$ of size $11 \times 11$, determined by 
the equations $ N_1 \, N_m = N_{m-1} + N_{m+1}$ ,   with $N_0 =  \one_{11}$ and $N_1$, the adjacency matrix of 
the diagram $A_{11}$.  One obtains in particular,  $N_1 \, N_{10} = 
N_9$ since $N_{11}$ vanishes.
Of course  $N_{m,n}^p = {(N_m)}_n^p$.

Using toric matrices, the general modular 
splitting equation reads therefore
$$
\sum_{m^{\prime \prime},n^{\prime \prime}}
\left(N_{m^{\prime \prime}}\right)_{m',m}
 \left(N_{n^{\prime \prime}}\right)_{n',n}
\left(W_{x,y}\right)_{m^{\prime \prime} n^{\prime \prime}} 
= 
\sum_z \left(W_{x,z}\right)_{mn}\left(W_{z,y}\right)_{m'n'}
$$ 
This  general equation, valid for any simply laced graph belonging to a Coxeter-Dynkin system, together with its proof and interpretation, in terms of associativity of
a bi-module structure,
was obtained in the 
thesis \cite{Gil:thesis}; it generalizes the ``modular splitting 
equation'' described in \cite{Ocneanu:Bariloche}. The later can be 
obtained from the former by setting $x=y=0$, so that its left hand 
side involves only known quantities, namely the matrix $M = W_{0,0}$
associated with the given graph
(in the simply laced cases, it is
{\sl the\/} modular invariant),  and the fusion 
coefficients.
It reads
$$
\sum_{m^{\prime \prime},n^{\prime \prime}}\left(N_{m^{\prime \prime}}\right)_{m',m}\left(N_{n^{\prime \prime}}\right)_{n',n}M_{m^{\prime \prime} n^{\prime \prime}} 
= \sum_z  \left(W_{z,0}\right)_{m'n'}  \left(W_{0,z}\right)_{mn}
$$
or, in terms of tensor products, 
$$\sum_{mn} N_m \otimes N_n \; M_{m,n} = \sum_{z \in Oc}  \tau \circ  (W_{z,0} \otimes 
W_{0,z})$$
Here $\tau$ denotes a tensorial flip (compare  explicit indices in the two previous equations). 
The left and side -- call it $K$ --  is therefore a known matrix of 
size $121 \times 121$ (in this case), with positive integer entries,  and the right 
hand side involves a set of toric matrices (with one twist), to be determined. 
Each term of this right hand side should be a matrix of rank $1$ with positive integer coefficients.  Only one member of 
this family is a priori known, namely $M=W_{0,0}$ which is our 
initial data. The indexing set on the right hand side of the modular splitting equation defines the set of vertices of the Ocneanu graph.
The problem at hand is analogous to those related with convex decompositions  in 
abelian mono\"{\i}ds.
In the simply laced cases (where  $W_{x,0}=W_{0,x}$ for all $x$),  each  term appearing on the right hand side is a tensor square composed with the flip.  In the non simply laced case, as we shall see, the situation is slightly more complicated. Notice that, in any case,  calling ``toric vectors'' $w_{x,y}$  the line-vectors obtained by ``flattening'' the toric matrices $W_{x,y}$ (\ie $ (w_{x,y})_k= W_{x,y}[p,q] $ with $k = (p-1)11 + q$), one can write the modular splitting equation as follows:
$K = \sum_z K_z$
where each matrix $K_z$ -- of size $121 \times 121$ with positive integral entries --  should be of rank $1$,  and its $k$-th line is equal to 
$$
K_z [k]  = (w_{z,0})_k \, w_{z,0} 
$$

\subsection*{Resolution of the equations}

The algorithm used to solve this set of equations (over the positive 
integers) is slightly different for the known simply-laced case and 
for the example that we study in this paper. 
Let us first summarize 
what we would do in the usual situation and call $n$ the total number 
of vertices of the corresponding fusion graph.

\subsubsection*{The simply laced case}

\begin{itemize}

\item The first line of $K$ -- a line vector with $n^2$ components 
--  is just the ``flattened'' invariant matrix $M$.
\item One does not assume that $W_{x,y}$ is equal to $W_{y,x}$ but 
takes nevertheless $W_{x} \doteq W_{x,0} = W_{0,x}$.
\item The rank $r(K)$ of $K$ can be calculated. This tells us that 
the dimension of the vector space spanned by the $n^2$ lines of $K$ 
can be expanded on a set of $r(K)$ basis vectors $w_x = w_{x,0}$ (the
toric vectors). Once a toric vector $w$ is found, the corresponding toric matrix 
is obtained by partitioning its entries into $n$  lines of length $n$.
\item The problem is to determine whether a given line of $K$ is a toric vector 
or if it is a positive integral linear 
combination of such vectors, and in that later case, one has to find 
the number of such terms in the sum. 
We choose the (non canonical) scalar product for which the basis of toric vectors is 
orthonormal;  for every line of $K$ we write $K[p] = \sum_x a(x) \, 
w_x $ and  the norm square of the vector $K[p]$ is therefore $\sum 
a(x)^2$.
The fundamental observation is that  this number (call it $\ell[p]$) is equal to the 
diagonal matrix element $K[p,p]$.
\item 
Since $K$ is known, we first consider the list of values of $p$ for 
which $K[p,p]=1$. 
Since several  line vectors $K[p]$ and $K[p']$  may coincide for 
distinct values of $p,p'$, we actually build a restricted list. For 
$p$ in this list, every  line vector $K[p]$ is then automatically a 
toric vector. 
\item 
We next consider the list (actually a restricted list, as above) of 
values of $p$ for which $K[p,p]=2$.  The corresponding line vectors 
$K[p]$ should be the sum of two toric vectors, and there are three 
cases.
Either $K[p]$ is the sum of two already determined toric vectors, or 
it is the sum of an already determined toric vector and a new one, 
else it is equal to twice a new toric vector. For every value of $p$ 
belonging to the new restricted list,  it is enough to calculate the 
set of differences $K[p] - w_x$ were $w_x$ runs in the set of the 
already determined toric vectors, and impose that all the components 
of such  differences should be positive integers.
\item
The next step is to consider the set of values of $p$ for which 
$K[p,p]=3$, \etc  and generalize  the previous discussion in an 
straightforward way. The process  stops, ultimately,  since the rank of the system 
is finite. At the end, we obtain a set of $r(K)$ toric vectors which 
are either lines of $K$ of linear combinations of lines of $K$.
\item The integer $r(K)$ may be strictly smaller than the number  $o$ 
of vertices of the Ocneanu graph. This happens when distinct quantum 
symmetries $x$ are associated with the same toric matrix $W_x$. This 
is for example the case of the graph $D_4$ where the rank is $5$ but 
where $o =8$. A method to determine such multiplicities is to plug 
the results for $w_x$ (actually for the matrices $W_x$) back into the 
modular splitting equation. If there are no multiplicities, this 
equation is readily checked. If it does not hold, one has to 
introduce appropriate multiplicities in the right hand side 
(introduce unknown coefficients and solve). In the case of $E_6$, the rank 
is $12$, the final list of toric vectors is obtained from lines 
$1,2,3,10,11,12,13,14,21,22,23,4-10$ of $K$, the last one being equal 
to a difference of two lines\footnote{Using the notations of 
\cite{Coque:Qtetra} or \cite{CoqueGil:ADE}, these toric vectors correspond 
respectively to
$W_{00}, W_{01},W_{02}$, $W_{05},W_{40},W_{10}$, 
$W_{11},W_{21},W_{51}$, $W_{50},W_{20},W_{30}.$}, and the modular 
splitting equation holds ``on the nose'', so $o=12$ also.
In the case of $D_4$, the rank is $5$, the  list of toric vectors is 
$K[1],K[2],K[6],K[3]/2,K[7]/3$ but the modular splitting equation 
holds only by introducing multiplicities $2$ and $3$ for the last 
two\footnote{Using the notations of \cite{Ocneanu:paths} or \cite{CoqueGil:ADE}, they 
correspond respectively to $W_0, W_{1\epsilon},  W_1$, $ W_2 = W_2'$, 
$W_\epsilon = W_{2\epsilon} = W_{2\epsilon}'$},  the number of 
quantum symmetries $o$ is therefore $8$.

\end{itemize}

\subsubsection*{The case of $F_4$ (non simply laced)}

\begin{itemize}

\item As usual, the first line $K[1]$ of $K$ -- a line vector with $121$ 
components --  is just the ``flattened'' invariant matrix $M$.
\item The rank of $K$ is $20$.   We present the results according to the decomposition number $\ell[p]$ relative to the line vector $K[p]$ of $K$,  with $p$ running from $1$ to $121$. The twenty toric vectors can be taken as 
follows.
$\ell[p]=1$: $w[p] = K[p]$ with $p=1,2,3,10,11,12,13,14,21,22,23,24$.

$\ell[p]=2$: 
$w[4] = K[4] - w[10] = K[4]-K[10]$,
$w[15] = K[15] - w[21]=K[15] - K[21]$,
$w[25] = K[25]/2$,
$w[34] = K[34] - w[22] = K[34]-K[22]$,
$w[35] = K[35] - w[21]=K[35]-K[21]$.

$\ell[p]=3$:
$w[26] = (K[26] - w[24])/2 = (K[26]-K[24])/2$,
$w[36] = (K[36] - w[14])/2 = (K[36]-K[14])/2 $.

$\ell[p]=5$:
$w[37]=(K[37] - w[13] - w[15] - w[35])/2 = K[37] -K[13] - K[15] + 2 K[21] - K [35])/2 $
\item
We  re - label the toric vectors
 in such a way that the index $x$ of $w_x$ runs from $0$ to $19$.

\begin{eqnarray*}
w_0=w[1], w_1 = w[2], w_2 = w[3], w_3 = w[4], w_4=w[10], \\
w_5=w[11], w_6=w[12], w_7=w[13],w_8=w[14],w_9=w[15], \\
 w_{10}=w[21], w_{11}=w[22],w_{12}=w[23], w_{13}=w[24], w_{14}=w[25],
\\ w_{15}=w[26],w_{16}=w[34],w_{17} =w[35], w_{18}=w[36], w_{19}=w[37]
\end{eqnarray*}
 
 There is a one to one correspondence between toric matrices $W_{x,0}$ and the 
previously determined toric vectors $w_x$ which have $121$ components :  
just partition them into $11$ lines of $11$ elements. 
 At a later stage we shall write the generators $O_x$ of the algebra $Oc$
 in terms of tensor products, for this reason, the  following distinct notations appear in this paper:
$W_x = W_{x,0} = W_{\underline{ab^\prime}}$ whenever $O_x=a \otimesdot b$.  
We hope that no confusion should arise between the notations  $W_{\underline{ab^\prime}}$ and $W_{x,y}$.
 
\item
Explicitly the twenty toric matrices $W_x$ are as follows. 

{\tiny

$$
\begin{array}{rr}
W_{0\,}=\left(\begin{array}{ccccccccccc} 
 1 & . & . & . & . & . & 1 & . & . & . & . \\ 
 . & . & . & . & . & . & . & . & . & . & . \\ 
 . & . & . & . & . & . & . & . & . & . & . \\ 
 . & . & . & . & . & . & . & . & . & . & . \\ 
 . & . & . & . & 1 & . & . & . & . & . & 1 \\ 
 . & . & . & . & . & . & . & . & . & . & . \\ 
 1 & . & . & . & . & . & 1 & . & . & . & . \\ 
 . & . & . & . & . & . & . & . & . & . & . \\ 
 . & . & . & . & . & . & . & . & . & . & . \\ 
 . & . & . & . & . & . & . & . & . & . & . \\ 
 . & . & . & . & 1 & . & . & . & . & . & 1 \\
\end{array}\right)&
 W_{5\,}=\left(\begin{array}{ccccccccccc}
. & . & . & . & 1 & . & . & . & . & . & 1 \\ 
. & . & . & . & . & . & . & . & . & . & . \\ 
. & . & . & . & . & . & . & . & . & . & . \\ 
. & . & . & . & . & . & . & . & . & . & . \\ 
1 & . & . & . & . & . & 1 & . & . & . & . \\ 
. & . & . & . & . & . & . & . & . & . & . \\ 
. & . & . & . & 1 & . & . & . & . & . & 1 \\ 
. & . & . & . & . & . & . & . & . & . & . \\ 
. & . & . & . & . & . & . & . & . & . & . \\ 
. & . & . & . & . & . & . & . & . & . & . \\ 
1 & . & . & . & . & . & 1 & . & . & . & . \\
\end{array}\right) \\
&\\
 W_{7\,}=\left(\begin{array}{ccccccccccc}
. & . & . & . & . & . & . & . & . & . & . \\ 
. & 1 & . & . & . & 1 & . & 1 & . & . & . \\ 
. & . & . & . & . & . & . & . & . & . & . \\ 
. & . & . & 1 & . & 1 & . & . & . & 1 & . \\ 
. & . & . & . & . & . & . & . & . & . & . \\ 
. & 1 & . & 1 & . & 2 & . & 1 & . & 1 & . \\ 
. & . & . & . & . & . & . & . & . & . & . \\ 
. & 1 & . & . & . & 1 & . & 1 & . & . & . \\ 
. & . & . & . & . & . & . & . & . & . & . \\ 
. & . & . & 1 & . & 1 & . & . & . & 1 & . \\ 
. & . & . & . & . & . & . & . & . & . & . \\
\end{array}\right)&
 W_{10}=\left(\begin{array}{ccccccccccc} 
. & . & . & . & . & . & . & . & . & . & . \\ 
. & . & . & 1 & . & 1 & . & . & . & 1 & . \\ 
. & . & . & . & . & . & . & . & . & . & . \\ 
. & 1 & . & . & . & 1 & . & 1 & . & . & . \\ 
. & . & . & . & . & . & . & . & . & . & . \\ 
. & 1 & . & 1 & . & 2 & . & 1 & . & 1 & . \\ 
. & . & . & . & . & . & . & . & . & . & . \\ 
. & . & . & 1 & . & 1 & . & . & . & 1 & . \\ 
. & . & . & . & . & . & . & . & . & . & . \\ 
. & 1 & . & . & . & 1 & . & 1 & . & . & . \\ 
. & . & . & . & . & . & . & . & . & . & . \\
\end{array}\right)\\
&\\
 W_{14}=\left(\begin{array}{ccccccccccc}
. & . & . & . & . & . & . & . & . & . & . \\ 
. & . & . & . & . & . & . & . & . & . & . \\
. & . & 1 & . & 1 & . & 1 & . & 1 & . & . \\ 
. & . & . & . & . & . & . & . & . & . & . \\ 
. & . & 1 & . & 1 & . & 1 & . & 1 & . & . \\ 
 & . & . & . & . & . & . & . & . & . & . \\ 
. & . & 1 & . & 1 & . & 1 & . & 1 & . & . \\ 
. & . & . & . & . & . & . & . & . & . & . \\ 
. & . & 1 & . & 1 & . & 1 & . & 1 & . & . \\ 
. & . & . & . & . & . & . & . & . & . & . \\ 
. & . & . & . & . & . & . & . & . & . & . \\
\end{array}\right)&
W_{19}=\left(\begin{array}{ccccccccccc}  
. & . & . & . & . & . & . & . & . & . & . \\ 
. & . & . & . & . & . & . & . & . & . & . \\ 
. & . & . & . & . & . & . & . & . & . & . \\ 
. & . & . & 1 & . & . & . & 1 & . & . & . \\ 
. & . & . & . & . & . & . & . & . & . & . \\ 
. & . & . & . & . & . & . & . & . & . & . \\ 
. & . & . & . & . & . & . & . & . & . & . \\ 
. & . & . & 1 & . & . & . & 1 & . & . & . \\ 
. & . & . & . & . & . & . & . & . & . & . \\ 
. & . & . & . & . & . & . & . & . & . & . \\ 
. & . & . & . & . & . & . & . & . & . & . \\
 \end{array}\right)\\
&\\
W_{1}=(W_{6\;})^T=\left(\begin{array}{ccccccccccc}
. & 1 & . & . & . & 1 & . & 1 & . & . & . \\ 
. & . & . & . & . & . & . & . & . & . & . \\ 
. & . & . & . & . & . & . & . & . & . & . \\ 
. & . & . & . & . & . & . & . & . & . & . \\ 
. & . & . & 1 & . & 1 & . & . & . & 1 & . \\ 
. & . & . & . & . & . & . & . & . & . & . \\ 
. & 1 & . & . & . & 1 & . & 1 & . & . & . \\ 
. & . & . & . & . & . & . & . & . & . & . \\ 
. & . & . & . & . & . & . & . & . & . & . \\ 
. & . & . & . & . & . & . & . & . & . & . \\ 
. & . & . & 1 & . & 1 & . & . & . & 1 & . \\
\end{array}\right)&
 W_{2}=(W_{12})^T=\left(\begin{array}{ccccccccccc}
. & . & 1 & . & 1 & . & 1 & . & 1 & . & . \\ 
. & . & . & . & . & . & . & . & . & . & . \\ 
. & . & . & . & . & . & . & . & . & . & . \\ 
. & . & . & . & . & . & . & . & . & . & . \\ 
. & . & 1 & . & 1 & . & 1 & . & 1 & . & . \\ 
. & . & . & . & . & . & . & . & . & . & . \\ 
. & . & 1 & . & 1 & . & 1 & . & 1 & . & . \\ 
. & . & . & . & . & . & . & . & . & . & . \\ 
. & . & . & . & . & . & . & . & . & . & . \\ 
. & . & . & . & . & . & . & . & . & . & . \\ 
. & . & 1 & . & 1 & . & 1 & . & 1 & . & . \\
\end{array}\right) \\
&\\
 W_{3}=(W_{16})^T=\left(\begin{array}{ccccccccccc}
. & . & . & 1 & . & . & . & 1 & . & . & . \\ 
. & . & . & . & . & . & . & . & . & . & . \\ 
. & . & . & . & . & . & . & . & . & . & . \\ 
. & . & . & . & . & . & . & . & . & . & . \\ 
. & . & . & 1 & . & . & . & 1 & . & . & . \\ 
. & . & . & . & . & . & . & . & . & . & . \\ 
. & . & . & 1 & . & . & . & 1 & . & . & . \\ 
. & . & . & . & . & . & . & . & . & . & . \\ 
. & . & . & . & . & . & . & . & . & . & . \\ 
. & . & . & . & . & . & . & . & . & . & . \\ 
. & . & . & 1 & . & . & . & 1 & . & . & . \\
\end{array}\right)&
W_{4}=(W_{11})^T=\left(\begin{array}{ccccccccccc}
. & . & . & 1 & . & 1 & . & . & . & 1 & . \\ 
. & . & . & . & . & . & . & . & . & . & . \\ 
. & . & . & . & . & . & . & . & . & . & . \\ 
. & . & . & . & . & . & . & . & . & . & . \\ 
. & 1 & . & . & . & 1 & . & 1 & . & . & . \\ 
. & . & . & . & . & . & . & . & . & . & . \\ 
. & . & . & 1 & . & 1 & . & . & . & 1 & . \\ 
. & . & . & . & . & . & . & . & . & . & . \\ 
. & . & . & . & . & . & . & . & . & . & . \\ 
. & . & . & . & . & . & . & . & . & . & . \\ 
. & 1 & . & . & . & 1 & . & 1 & . & . & . \\
\end{array}\right)\\
&\\
W_{8}=(W_{13})^T=\left(\begin{array}{ccccccccccc}
. & . & . & . & . & . & . & . & . & . & . \\
. & . & 1 & . & 1 & . & 1 & . & 1 & . & . \\ 
. & . & . & . & . & . & . & . & . & . & . \\ 
. & . & 1 & . & 1 & . & 1 & . & 1 & . & . \\ 
. & . & . & . & . & . & . & . & . & . & . \\ 
. & . & 2 & . & 2 & . & 2 & . & 2 & . & . \\ 
. & . & . & . & . & . & . & . & . & . & . \\ 
. & . & 1 & . & 1 & . & 1 & . & 1 & . & . \\ 
. & . & . & . & . & . & . & . & . & . & . \\ 
. & . & 1 & . & 1 & . & 1 & . & 1 & . & . \\ 
. & . & . & . & . & . & . & . & . & . & . \\
\end{array}\right)&
W_{9}=(W_{17})^T=\left(\begin{array}{ccccccccccc}
. & . & . & . & . & . & . & . & . & . & . \\ 
. & . & . & 1 & . & . & . & 1 & . & . & . \\ 
. & . & . & . & . & . & . & . & . & . & . \\ 
. & . & . & 1 & . & . & . & 1 & . & . & . \\ 
. & . & . & . & . & . & . & . & . & . & . \\ 
. & . & . & 2 & . & . & . & 2 & . & . & . \\ 
. & . & . & . & . & . & . & . & . & . & . \\ 
. & . & . & 1 & . & . & . & 1 & . & . & . \\ 
. & . & . & . & . & . & . & . & . & . & . \\ 
. & . & . & 1 & . & . & . & 1 & . & . & . \\ 
. & . & . & . & . & . & . & . & . & . & . \\
\end{array}\right)\\
\end{array}
$$
$$
 W_{15}=(W_{18})^T=\left(\begin{array}{ccccccccccc} 
. & . & . & . & . & . & . & . & . & . & . \\ 
. & . & . & . & . & . & . & . & . & . & . \\ 
. & . & . & 1 & . & . & . & 1 & . & . & . \\ 
. & . & . & . & . & . & . & . & . & . & . \\ 
. & . & . & 1 & . & . & . & 1 & . & . & . \\ 
. & . & . & . & . & . & . & . & . & . & . \\ 
. & . & . & 1 & . & . & . & 1 & . & . & . \\ 
. & . & . & . & . & . & . & . & . & . & . \\ 
. & . & . & 1 & . & . & . & 1 & . & . & . \\ 
. & . & . & . & . & . & . & . & . & . & . \\ 
. & . & . & . & . & . & . & . & . & . & . \\
\end{array}\right)\\
$$
}

Notice that ${W_0 , W_5 , W_7 , W_{10} , W_{14} , W_{19}}$ are 
symmetric (self-dual). Among them, two will be called ``ambichiral'', namely 
$W_0$ and $W_5$ for the reason that they correspond to the ambichiral
generators of the algebra $Oc$ (intersection of the two chiral sub-algebras).

\item As expected from the presence of the $1/2$ coefficients in the 
 list giving the vectors $w[p]$, the modular splitting equation does not hold if we 
impose $W_{x,0} = W_{0,x}$ and sum only on the corresponding $20$ terms on the right 
hand side.
 One possibility is two introduce a multiplicity two for  entries 
$w_{14}=w[25] , w_{15}=w[26], w_{18}=w[36], w_{19}=w[37]$;  this indeed works, in the sense 
that the modular splitting equation is then satisfied.  With such a 
choice, the number of quantum symmetries would be $o = 24$, rather 
than $20$. However, the algebra of quantum symmetries later 
determined by this choice incorporates 
several arbitrary coefficients that cannot be fixed by requirements 
of positivity and integrality alone.
 Since we are in ``Terra Incognita'' (namely quantum symmetries of 
non simply laced diagrams), we prefer to explore another possibility, 
which also allows us to solve the modular splitting equation  and 
leads to a nice  algebra of quantum symmetries (and, as we shall see, 
to the emergence of the diagram $F_4$). Our choice is to keep only 
the previously determined  $20$ terms, no more, but without imposing 
equality of $W_{x,0}$ and $W_{0,x}$. This choice is natural in view of the fact that these
matrices actually ``count'' a number of essential paths between the origin $0$ and $x$
on the Ocneanu graph itself, and the fact that in the present case, the graph is not symmetric (all edges are not bi-oriented).
 In general, the solution to the modular splitting 
equation, for a given invariant matrix $M$  is not necessarily 
unique, although some later considerations may impose extra 
conditions that, ultimately, lead to rejection of one or another 
solution. At the moment, we investigate one solution 
(which is both minimal in terms of number of quantum symmetries and natural from the path interpretation point of view) and 
explore the consequences.

In order for the modular splitting equation to be satisfied, we therefore
take $W_{0,x} = 2 W_{x,0}$ when $x = 14, 15, 18, 19$ and  $W_{0,x}=W_{x,0}$
for the others.  As we shall see, this corresponds to the 
fact that the $F_4$ diagram contains an oriented edge.

\end{itemize}

\section{Quantum symmetries and Ocneanu graph} 

\subsection*{Determination of the two chiral generators $O_1$ and 
$O_1'$}

Call $K_0$ the rectangular matrix ($121 \times 20$) obtained by 
decomposing each line vectors of $K$ on the (flattened) toric 
matrices. For instance the fourth line $K[4]$ of $K$ is equal to $w[4] + 
w[5]$, so its components are 
$(0,0,0,1,1,0,0,0,0,0,0,0,0,0,0,0,0,0,0,0)$.

Call $L_0$ the rectangular matrix  obtained by transposing the matrix 
($20 \times 121$) obtained by flattening each component of the column vector (twenty lines)
containing the toric matrices.

When $W_{x,0}=W_{0,x}$ it is easy to see that $K_0 = L_0$ but this 
is not so in the present case.

If $E_0$ denotes the essential matrix  (also called ``intertwiner'') 
associated with the origin of an $ADE$ diagram (for the $SU(2)$ system 
or higher generalizations), and if $G_1$ denotes the corresponding 
adjacency matrix, it is so
that $E_1 \doteq N_1 \,  . \, E_0 =  E_0 \,  . \,  G_1$ where $N_1$ is 
the generator of the fusion algebra (adjacency matrix of the 
appropriate $A_n$ diagram). $E_1$ coincides with the essential matrix associated 
with the next vertex (after the origin) and describes essential paths emanating from it.

We have the following analogy:  $K_0$ (or $L_0$) play the same role as $E_0$, but now $G_1$ 
should be replaced by one of the two generators of the algebra of quantum symmetries, 
and $N_1$ should be replaced by $N_0 \otimes N_1$ (so we replace the 
fusion algebra by its tensor square). In other words, we determine 
the generator $O_1$  by solving 
the intertwining equation 
$$ N_0 \otimes N_1  \, .   \, K_0 = K_0 \, . \, O_1$$

The other chiral generator $O_1^\prime$ is determined by solving the same 
equation, but replacing $N_0 \otimes N_1$ by $N_1 \otimes N_0$.
At this level it is interesting to recall that, in this analogy
between the vector space of a diagram and its algebra of quantum 
symmetries, the fusion algebra should be replaced by its tensor 
square, and the role of fused adjacency matrices (the $F_{ab}$ 
matrices of references \cite{CoqueGil:ADE} that represent the action of $A_n$ on 
a given diagram)  is played by the toric matrices themselves.

In the present situation (non simply laced), we could hesitate 
between $L_0$ and $K_0$, but the choice actually does not 
matter :  it turns out that this arbitrariness  corresponds to 
the arbitrariness in the association between the asymmetric $F_4$ graph 
and a particular adjacency matrix or its transpose.

In general, after having solved the modular splitting 
equation (and there is no necessarily uniqueness of the solution), we 
have to solve a generalized intertwining identity (the one just 
given) in order to find the two chiral generators. Notice that the 
solution could be non unique, even after imposing integrality and 
positivity, but in the present case the solution is unique and we list\footnote{The reader already recognizes, from the structure of $O_1$, two subdiagrams of type $E_6$ and two others of type $F_4$} 
below the two matrices $O_1$ and $O_1^\prime$,  of dimension $20 \times 20$,
which solve these equations.

{\tiny
$$O_1 = \left(\begin{array}{cccccccccccccccccccc}

. & 1 & . & . & . & . & . & . & . & . & . & . & . & . & . & . & . & . 
& . & . \cr 
1 & . & 1 & . & . & . & . & . & . & . & . & . & . & . & . & . & . & . 
& . & . \cr 
. & 1 & . & 1 & 1 & . & . & . & . & . & . & . & . & . & . & . & . & . 
& . & . \cr 
. & . & 1 & . & . & . & . & . & . & . & . & . & . & . & . & . & . & . 
& . & . \cr 
. & . & 1 & . & . & 1 & . & . & . & . & . & . & . & . & . & . & . & . 
& . & . \cr 
. & . & . & . & 1 & . & . & . & . & . & . & . & . & . & . & . & . & . 
& . & . \cr 
. & . & . & . & . & . & . & 1 & . & . & . & . & . & . & . & . & . & . 
& . & . \cr 
. & . & . & . & . & . & 1 & . & 1 & . & . & . & . & . & . & . & . & . 
& . & . \cr 
. & . & . & . & . & . & . & 1 & . & 1 & 1 & . & . & . & . & . & . & . 
& . & . \cr 
. & . & . & . & . & . & . & . & 1 & . & . & . & . & . & . & . & . & . 
& . & . \cr 
. & . & . & . & . & . & . & . & 1 & . & . & 1 & . & . & . & . & . & . 
& . & . \cr 
. & . & . & . & . & . & . & . & . & . & 1 & . & . & . & . & . & . & . 
& . & . \cr 
. & . & . & . & . & . & . & . & . & . & . & . & . & 1 & . & . & . & . 
& . & . \cr 
. & . & . & . & . & . & . & . & . & . & . & . & 1 & . & 2 & . & . & . 
& . & . \cr 
. & . & . & . & . & . & . & . & . & . & . & . & . & 1 & . & 1 & . & . 
& . & . \cr 
. & . & . & . & . & . & . & . & . & . & . & . & . & . & 1 & . & . & . 
& . & . \cr 
. & . & . & . & . & . & . & . & . & . & . & . & . & . & . & . & . & 1 
& . & . \cr 
. & . & . & . & . & . & . & . & . & . & . & . & . & . & . & . & 1 & . 
& 2 & . \cr 
. & . & . & . & . & . & . & . & . & . & . & . & . & . & . & . & . & 1 
& . & 1 \cr 
. & . & . & . & . & . & . & . & . & . & . & . & . & . & . & . & . & . 
& 1 & . \cr
\end{array}\right)
$$

$$
O'_1 =\left(\begin{array}{cccccccccccccccccccc}

. & . & . & . & . & . & 1 & . & . & . & . & . & . & . & . & . & . & . 
& . & . \\ 
. & . & . & . & . & . & . & 1 & . & . & . & . & . & . & . & . & . & . 
& . & . \\ 
. & . & . & . & . & . & . & . & 1 & . & . & . & . & . & . & . & . & . 
& . & . \\ 
. & . & . & . & . & . & . & . & . & 1 & . & . & . & . & . & . & . & . 
& . & . \\ 
. & . & . & . & . & . & . & . & . & . & 1 & . & . & . & . & . & . & . 
& . & . \\ 
. & . & . & . & . & . & . & . & . & . & . & 1 & . & . & . & . & . & . 
& . & . \\ 
1 & . & . & . & . & . & . & . & . & . & . & . & 1 & . & . & . & . & . 
& . & . \\ 
. & 1 & . & . & . & . & . & . & . & . & . & . & . & 1 & . & . & . & . 
& . & . \\ 
. & . & 1 & . & . & . & . & . & . & . & . & . & . & . & 2 & . & . & . 
& . & . \\ 
. & . & . & 1 & . & . & . & . & . & . & . & . & . & . & . & 2 & . & . 
& . & . \\ 
. & . & . & . & 1 & . & . & . & . & . & . & . & . & 1 & . & . & . & . 
& . & . \\ 
. & . & . & . & . & 1 & . & . & . & . & . & . & 1 & . & . & . & . & . 
& . & . \\ 
. & . & . & . & . & . & 1 & . & . & . & . & 1 & . & . & . & . & 1 & . 
& . & . \\ 
. & . & . & . & . & . & . & 1 & . & . & 1 & . & . & . & . & . & . & 1 
& . & . \\ 
. & . & . & . & . & . & . & . & 1 & . & . & . & . & . & . & . & . & . 
& 1 & . \\ 
. & . & . & . & . & . & . & . & . & 1 & . & . & . & . & . & . & . & . 
& . & 1 \\ 
. & . & . & . & . & . & . & . & . & . & . & . & 1 & . & . & . & . & . 
& . & . \\ 
. & . & . & . & . & . & . & . & . & . & . & . & . & 1 & . & . & . & . 
& . & . \\ 
. & . & . & . & . & . & . & . & . & . & . & . & . & . & 1 & . & . & . 
& . & . \\ 
. & . & . & . & . & . & . & . & . & . & . & . & . & . & . & 1 & . & . 
& . & . \\
\end{array}\right)
$$
}

Now that we have obtained the two algebraic  generators, the algebra $Oc$
that they span is determined as well (take linear combinations of 
powers and products). However  we shall exhibit a 
particular basis of linear generators.
To find them, one possibility is to first determine the set of 
matrices $V_{m,n}$ that describe, in a dual way, the full set of 
toric matrices $W_{x,y}$ with two twists.

\subsection*{Determination of the toric matrices with two twists}

The $V_{m,n}$ are obtained by solving the general intertwining equation (notice that $O_1 
= V_{0,1}$ and $O_1^\prime = V_{1,0}$).
$$ N_m \otimes N_n   \, . \,  K_0 = K_0  \, . \,  V_{m,n}$$ 
The solution is unique. Of course, we shall not list these $11^2$ square matrices of 
dimension $12 \times 12$ and we shall not give, either, the list of toric matrices with two twists, but
remember that they are determined by the 
relation
$$ {(W_{x,y})}_{m,n} = {(V_{m,n})}_{x,y} $$
One can then check that the generalized equation of modular splitting 
(the one that involves the $W_{x,y}$ rather than the $W_{x,0}$)  is 
satisfied.

\subsection*{Determination of the linear  generators  $O_x$ of $Oc$}

The structure constants $O_{xyz}$ are defined by   the equations
$$ W_{y,x} = \sum_z O_{xyz} \, W_z$$
where $W_z = W_{z,0}$.  Notice that, in the present case,   $O_{xyz}=O_{zyx}$ but $O_{xyz} \neq O_{yxz}$ in general. Matrices $O_x$ are defined by their  
coefficients as follows : $ (O_x)_{yz} = O_{xyz}$. We have $$O_x \, O_y = \sum_z O_{xzy} \, O_z$$ and any two generators $O_x$ and $O_y$ commute,  because of the symmetry properties of the structure constants.

One could be tempted to consider the (non-commutative) family of matrices $Z_y$ defined by the equation  $ (Z_y)_{xz} = O_{xyz}$ but one can see that this family is not multiplicatively closed; moreover $Z_0$ does not even co\" \i ncide with the identity matrix, since it has diagonal coefficients equal to $2$ in positions $x = 14, 15, 18, 19$.

When the Ocneanu diagram possesses geometric symmetries, for instance 
in the case of the $D$ diagrams, it may be that the general solution involve parameters that should be fixed by 
imposing positivity and integrality, and that one solution 
 is only determined up to a discrete transformation reflecting the classical symmetries (this amounts to re-label the vertices $x$). In the present case, however, everything is perfectly 
determined and we obtain the twenty  generators of the 
Ocneanu diagram -- They are $20 \times 20$ matrices. Rather than 
giving this list explicitly (it would be typographically heavy !), we 
shall express them in terms of the already known and explicitly 
given chiral generators $O_1$ and $O_1^\prime$.  Call $O_0$ the unit 
matrix  $\one_{20}$.

{\footnotesize
 $$
 \begin{array}{|lc|lc|lc|l}
&&&&&&\\
   O_0                 &&O_6=O_1^\prime            &&O_{12}=O_6.O'_1-O_0&&O_{16}=O_{12}.O'_1-O_6-O_{11} \\
  O_1                  &&O_7=O_1\cdot O'_1   &&\downarrow &&\downarrow\\
O_2=O_1^2-O_0          &&O_8=O_2\cdot O'_1   &&O_{13}=O_{12}\cdot O_1&&O_{17}=O_{16}\cdot O_1\\
O_5=O_1^4-4 O_1^2+2 O_0&&O_9=O_3\cdot O'_1   &&O_{14}=O_{13}\cdot O_1-O_{12}&&O_{18}=O_{17}\cdot O_1-O_{16}\\
  O_4=O_1\cdot O_5     &&O_{10}=O_4\cdot O'_1&&O_{15}=O_{14}\cdot O_1-2 O_{13}&&O_{19}=O_{18}\cdot O_1-2 O_{17}\\
  O_3=O_2.O_1-O_4-O_1  &&O_{11}=O_5\cdot O'_1&& O_{14}=O_{15}\cdot O_1 (check)  &&O_{18}=O_{19}\cdot O_1 (check) \\
\end{array}
$$
}

The full multiplication table (that we don't display  because  it is $20\times 20$) defines a 20 dimensional 
algebra $Oc$ with linear generators  $O_x$,  with $x \in \{0,1,2,3,\ldots 18,19\}$.
It is generated, as an algebra, by the two matrices associated with vertices $1$ and $6$ (called
{\sl chiral generators\/}).
 We call it  ``algebra of quantum symmetries of $F_4$''.  Multiplication of any single linear generator $O_x$ by the two chiral ones  is encoded by a graph: the Ocneanu graph of $F_4$.  It will be described later.

\subsection*{The Ocneanu algebra as a the tensor square of a graph 
algebra}

For all simply laced diagrams belonging to the $SU(2)$ system or to 
an higher system,  the algebra of quantum symmetries turns out to be
related, in one way or another, to the tensor square of some graph 
algebra. For instance $Oc(E_{6})$ is the tensor square of the graph
algebra of $E_{6}$ taken above the graph subalgebra generated by the 
ambichiral vertices $0, 4, 3$. We remind the reader that $E_6$ admits self - fusion, with  graph algebra given by the following table. 

\begin{figure}[htb]
\unitlength 0.6mm
\begin{center}
$$
\begin{array}{lcr}
\begin{picture}(95,35)
\thinlines
\multiput(25,7)(15,0){5}{\circle*{2}}
\put(55,22){\circle*{2}}
\thicklines
\put(25,7){\line(1,0){60}}
\put(55,7){\line(0,1){15}}
\put(25,0){\makebox(0,0){$\underline 0$}}
\put(40,0){\makebox(0,0){$\underline 1$}}
\put(55,0){\makebox(0,0){$\underline 2$}}
\put(70,0){\makebox(0,0){$\underline 5$}}
\put(85,0){\makebox(0,0){$\underline 4$}}
\put(63,24){\makebox(0,0){$\underline 3$}}
\end{picture}&\longleftrightarrow&
 \begin{array}{||c||c|c|c||c|c|c||}
 \hline
 * & \underline  0 & \underline 3  &\underline 4 &\underline 1 &\underline 2 &\underline 5   \\
 \hline
 \hline
\underline 0 &\underline 0 &\underline 3 &\underline 4  &\underline 1 &\underline 2 &\underline 5 \\
\underline 3 &\underline 3 &\underline 0+\underline 4  &\underline 3 &\underline 2 &\underline 1+\underline 5 &\underline 2 \\
\underline 4 &\underline 4 &\underline 3 &\underline 0 &\underline 5 &\underline 2 &\underline 1 \\
 \hline
\underline 1 &\underline 1 &\underline 2 &\underline 5 &\underline 0+\underline 2 &\underline 1+\underline 3+\underline 5 &\underline 2+\underline 4   \\
\underline 2 &\underline 2 &\underline 1+\underline 5 &\underline 2 &\underline 1+\underline 3+\underline  5 &\underline 0+\underline 2+\underline 2+\underline 4 &\underline 1+\underline 3+\underline 5 \\
\underline 5&\underline 5  &\underline 2 &\underline 1 &\underline 2+\underline 4 &\underline 1+\underline 3+\underline 5 &\underline 0+\underline 2 \\
 \hline
 \end{array}
\end{array}
$$
\label{graphE6}
\end{center}
\end{figure}

 It is natural to try to 
realize $Oc(F_{4})$, that we obtained by solving the modular 
splitting equation, directly in terms of some analogous algebraic construction. From
the fact that $F_{4}$ is an orbifold of $E_{6}$, it is  easy to make an educated
guess, and, by calculating the corresponding multiplication table, check that it is indeed correct.
We claim that $Oc(F_{4}) = E_{6} \otimesdot E_{6}$ where $\otimesdot$ 
denotes the tensor product taken, not above the complex numbers but above 
the subalgebra $J$ generated by vertices $ \underline 0$ and $\underline 4$ of $E_6$. In other words, we identify $a*u \otimes b$ and $a \otimes u*b$ as soon as $u \in J$.

The twenty generators of $Oc(F_{4})$ are realized as follows.

{\small
\centerline{
$
\begin{array}{ccc}
{} & \underline 0 = \underline 0 \otimesdot 0 = \underline 4 \otimesdot \underline 4 = \underline 0& {} \\
1 = \underline 1  \otimesdot \underline 0 = \underline 5  \otimesdot \underline 4 =\underline 1 & {} &6 = \underline 0  \otimesdot\underline 1 = \underline 4  \otimesdot \underline 5 =  \underline 1^\prime \\
2 = \underline 2  \otimesdot \underline 0 = \underline 2  \otimesdot \underline 4 =\underline 2 & {} &12= \underline 0  \otimesdot \underline 2 = \underline 4  \otimesdot \underline 2 =  \underline 2^\prime \\
3 = \underline 3  \otimesdot \underline 0 = \underline 3  \otimesdot \underline 4 =\underline 3 & {} &16= \underline 0  \otimesdot \underline 3 = \underline 4  \otimesdot \underline 3 =  \underline 3^\prime \\
4 = \underline 5  \otimesdot \underline 0 = \underline 1  \otimesdot \underline 4 =\underline 5 & {} &11= \underline 0  \otimesdot \underline 5 = \underline 4  \otimesdot \underline 1 = \underline  5^\prime \\
{} &5 = \underline 4  \otimesdot \underline 0 = \underline 0  \otimesdot \underline 4 = \underline 4 & { } \\
{} &7 = \underline 1  \otimesdot \underline 1 = \underline 5  \otimesdot \underline 5 =  \underline{11^\prime} & {} \\
{} &10= \underline 5  \otimesdot \underline 1 = \underline 1
\otimesdot \underline 5 =  \underline{15^\prime} & {} \\
8 = \underline 2  \otimesdot \underline 1 = \underline 2 \otimesdot
\underline 5 = \underline{21^\prime} &{}& 13 = \underline 1\otimesdot
\underline 2= \underline 5\otimesdot \underline 2=\underline{12^\prime} \\
{} &14 = \underline 2  \otimesdot \underline 2 =  \underline{22^\prime} & {} \\
9 = \underline 3\otimesdot \underline 1 = \underline 3\otimesdot \underline 5 =\underline{ 31^\prime} &{}&17 = \underline 1\otimesdot \underline 3 = \underline 5\otimesdot \underline 3 =\underline{ 13^\prime} \\
15= \underline 3\otimesdot \underline 2 = \underline{32^\prime} & {} & 18 = \underline 2  \otimesdot \underline 3 =  \underline{23^\prime} \\
{} &19 = \underline 3  \otimesdot \underline 3 =  \underline{33^\prime} & {}
\end{array}
$
}}

 Labels on the left correspond to the original notation $\{0,1,2\dots 19\}$ that we have been using before, 
while underlined  labels on the right refer to tensor products of $E_6$ vertices (as defined\footnote{Warning: There are several conventions in the literature} by the above $E_6$ diagram, like in \cite{Ocneanu:paths}, \cite{Coque:Qtetra} or \cite{Gil:thesis}). 
Because of this labeling convention, notice that $4 = \overline 5$ and $5 = \overline 4$.
Using the above realization, one recovers the multiplication table 
of quantum symmetries. For instance, $\underline{21^\prime} \times \underline{23^\prime}  =( \underline{22^\prime})_2 + (\underline{2^\prime})_2  $. Indeed,
$$ 
8 \times  18 = ( \underline 2  \otimesdot \underline 1)\times(
\underline 2  \otimesdot \underline 3) = ( \underline 2 * \underline
2)  \otimesdot  ( \underline 1 * \underline 3) = ( \underline 0 +
\underline 2 + \underline 2 + \underline 5) \otimesdot ( \underline 2)
= ( \underline 0 \otimes \underline 2)_2 + ( \underline 2 \otimes \underline 2)_2 =
12 + 12+ 14 + 14 
$$
We therefore recover the matrix product equality
$
O_{8} \times  O_{18} = 2 O_{12} + 2 O_{14}
$

\subsection*{The Ocneanu graph}

Using $E_{6} \otimesdot E_{6}$ notation for the vertices, the $F_4$ Ocneanu graph is given as follows:

\centerline{\includegraphics[width=100mm,height=100mm]{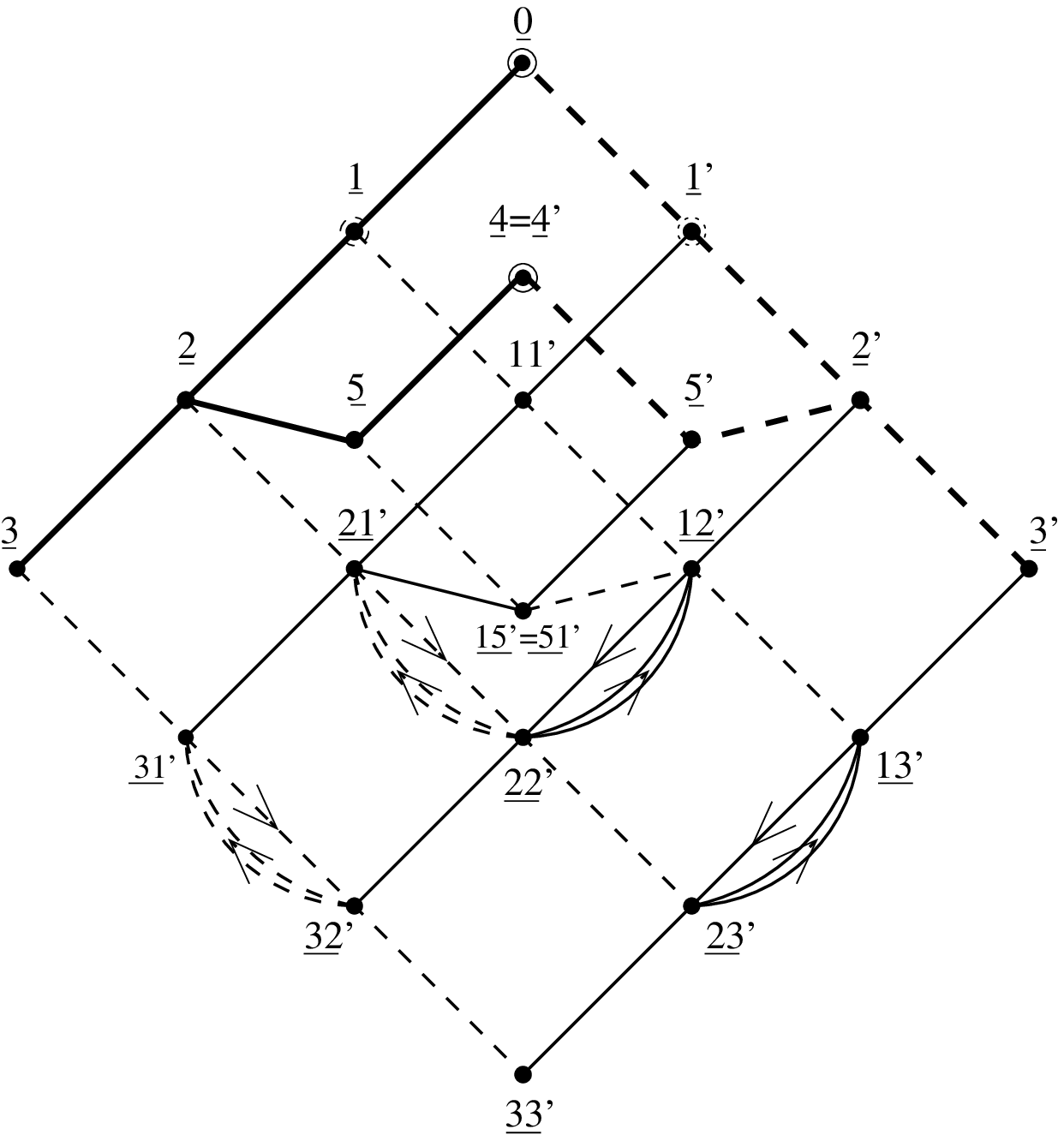}}

It is the Cayley graph of multiplication of the linear generators of $Oc$ by the two generators
$O_{1}$ and $O_{1}'$, called the {\sl chiral\/} generators. It is 
the union of two distinct graphs called left and right graphs, 
involving the same set of vertices. They are drawn in two different colors (or solid and dashed lines). 
One can obtain one graph from the other by performing a symmetry with 
respect to the vertical axis. The six vertices that belong to this axis of 
symmetry are the {\sl self-dual points}.

The subalgebra generated by the unit and the left chiral generator $\underline 1$  is the left chiral subalgebra. It is spanned by vertices $\{\underline 0, \underline 1, \underline 2, \underline 5, \underline 4, \underline 3\}$ and corresponds to the connected component of the identity of the left chiral graph.
The subalgebra generated by the unit and the right chiral generator $\underline 1^\prime $ is the right chiral subalgebra. It is spanned by vertices $\{\underline 0, \underline 1^\prime, \underline 2^\prime,   \underline 5^\prime, \underline 4^\prime, \underline 3^\prime\}$ and corresponds to the connected component of the identity of the right chiral graph.
The intersection of these two subalgebras  is spanned by the set of {\sl ambichiral points}, 
namely the two-element set $\{\underline 0, \underline 4 \}$. These points are self-dual.

Because of these symmetry considerations, we discuss only the left  graph.
It is itself given by the union of four disjoint connected graphs, two of type $E_{6}$, 
that we call respectively $E=E_{6}[1]$ (the left chiral subalgebra) and $e=E_{6}[2]$ (span of $\underline{1^\prime},\underline{11^\prime},\underline{21^\prime},\underline {51^\prime}, \underline{5^\prime},\underline{31^\prime}$),  and two of type $F_{4}$, that 
 we call $F=F_{4}[1]$ (span of $\underline{32^\prime}, \underline{22^\prime}, \underline{12^\prime}, \underline{2^\prime}$) and $f=F_{4}[2]$ (span of $\underline{33^\prime},\underline{23^\prime},\underline{13^\prime},\underline 3^\prime$). The description of the right graph is similar.
Notice that  $F_{4}$ Dynkin diagrams  emerge from our resolution of the equations of modular 
splitting.

The first $E_{6}$ (left) subgraph called $E$ describes the subalgebra generated by $O_{1}$. The 
other $E_{6}$ (left) subgraph called $e$  is not a 
subalgebra of $Oc$ but a module over $E$.
The two subgraphs of  type $F_4$ called $F$ and $f$ are also modules over $E$, but their 
properties are very different. Writing the full multiplication table 
would be too long, but the interested reader can easily do it, either 
using   $20 \times 20$ matrices, or, more simply, using 
the multiplication table of the graph algebra of $E_{6}$ together with 
the realization of generators of $Oc(F_{4})$ in terms of tensor 
products (see previous section). We have the following  relations
between the different subspaces:
$$
\begin{array}{||c||c|c|c|c||}\hline
{\times} & E & e & F & f \\
\hline
\hline
E & E  & e & F & f \\
e & e & E + F & e + f  & F \\
F & F  & e+f  & E + F & e \\
f & f & F & e &  E\\ \hline
\end{array}
$$

\section{Actions, coactions and  sum rules}

As written in the introduction, the Ocneanu quantum groupoid associated with a simply laced diagram 
$G$ (with $r$ vertices) belonging to the $SU(2)$ system, or to an higher system, possesses 
two -- usually distinct -- algebras of characters.
The first,  called the fusion algebra of $G$, 
and denoted $A(G)$,  can be identified with the graph algebra of 
the graph $A_n$ (for a proper choice of $n$).
The second algebra of 
characters, called the algebra of quantum symmetries, is denoted 
$Oc(G)$.
The fusion algebra $A(G)=A_{n}$ acts on the vector space spanned by 
the vertices of the 
graph $G$, and this action is explicitly described by the so-called 
``fused adjacency matrices'' $F_{p}$. These matrices have therefore 
the same commutation relations as the fusion matrices $N_{p}$ (the 
generators of $A_{n}$) but their size is smaller since it is $r 
\times r$, where $r$ is the number of vertices of $G$.
In the same way, but at the dual level (coaction), the algebra of quantum symmetries $Oc(G)$
acts on  $G$ and this  is described by the 
so-called ``quantum symmetry fused matrices'' $\Sigma_{x}$ of $G$.  These matrices have therefore 
the same commutation relations as the quantum symmetry matrices $O_{x}$ (the 
generators of $Oc(G)$) but their size is smaller since it is again $r \times r$. See \cite{Coque:Qtetra}, \cite{CoqueGil:ADE} for explicit expressions of these matrices in various $ADE$ cases.
In the simply laced situation, there is a general theory \cite{Ocneanu:paths}  (see also
 \cite{CoTr-DTE}, \cite{PetZub:Oc}) 
that tells us how to build first, a product  law -- composition -- defined as composition
of graded endomorphisms of essential paths (``horizontal paths'') on 
the given graph, then a co-product law -- associated, via the choice of a scalar product, to convolution -- by using the composition of
endomorphisms of the so - called ``vertical paths''. However, to our knowledge, for a non simply-laced diagram 
like the one we study here, the general theory is not known, yet.
 Our purpose, in this section,
is therefore very modest, in the sense that we shall only mimic what we would have done in the
simply laced situation, and describe what we find. This is admittedly rather naive, since when counting dimensions,  for instance, we take the oriented double line of the $F_4$ diagram (between vertices $a_2$ and $a_1$) as a pair of two essential paths of length one, and this is maybe not what should be done.

\subsection*{Fused matrices $F_{p}$ relative to the fusion generators $N_p$ of $A_{11}$}

We first consider the action of $A_{11}$ implemented by matrices 
$F_{p}$, to be found. The simplest determination stems from the 
fact that $A_{11}$ is a truncated version of the algebra of 
characters of $SU(2)$, so that the $F_{p}$'s are obtained from the 
usual recurrence formula (composition of spins) $F_{p} F_{1} = 
F_{p-1} + F_{p+1}$, and the seed: $F_{0}=\one_{4}$ and $F_{1}$ is equal to the 
adjacency matrix $G_1$ of the graph $F_{4}$. This recurrence relation has a 
period ($2 \times 12$) and one can check that $F_{10} F_{1} = F_{9}$ 
since $F_{11}$ vanishes. For $12 \leq p \leq 23$, $F_{p+12}=-F_{p}$ but 
we are only interested here in the positive part. Notice that $F_{1}$ is 
not symmetric, since the graph $F_{4}$ itself, with vertices $a_0,a_1,a_2,a_3$,
is not\footnote{The reader may check that $C = 2 \, \one - G_1$ is the usual Cartan matrix of $F_4$} (two oriented edges between vertices $a_2$ and $a_1$ but only 
one oriented edge between $a_1$ and $a_2$).   We obtain the following 
$11$ matrices and check that $(F_p      F_q) \,  a_i = F_p \, (F_q \,  a_i) $, as it should.
{\tiny
$$
\begin{array}{cccccc}
F_0 & F_1 & F_2 & F_3 & F_4 & F_5 \\
{} \\
\left(\begin{array}{cccc}
1&.&.&. \\
.&1&.&. \\
.&.&1&. \\
.&.&.&1 
\end{array}\right)
 &
\left(\begin{array}{cccc}
.&1&.&. \\
1&.&2&. \\
.&1&.&1 \\
.&.&1&. 
\end{array}\right)
&
\left(\begin{array}{cccc}
.&.&2&. \\
.&2&.&2 \\
1&.&2&. \\
.&1&.&. 
\end{array}\right)
&
\left(\begin{array}{cccc}
.&1&.&2 \\
1&.&4&. \\
.&2&.&1 \\
1&.&1&. 
\end{array}\right)
&
\left(\begin{array}{cccc}
1&.&2&. \\
.&3&.&2 \\
1&.&3&. \\
.&1&.&1 
\end{array}\right)
&
\left(\begin{array}{cccc}
.&2&.&. \\
2&.&4&. \\
.&2&.&2 \\
.&.&2&. 
\end{array}\right)
\end{array}
$$
}
and $ F_p = F_{10-p}$ for $p=6,\ldots, 10$.
 From the fused adjacency matrices $F_p$ we can obtain  four essential matrices $E_{a}$ that are 
rectangular $4\times 11$ matrices defined by 
$(E_{a})_{b,n}=(F_{n})_{a,b}$.
The $F$'s 
and the $E$'s determine the induction/restriction rules between the 
graphs $A_{11}$ and $F_{4}$.

\subsection*{Fused matrices $\Sigma_x$ relative to the quantum symmetry generators   $O_x$ of $Oc$}

We now turn to the determination of matrices $\Sigma_x$.
We shall present two methods. The most direct uses the fact that these matrices provide
a $4 \times 4$ realization of $Oc$. It is enough to define $\Sigma_0 = \one_4$,  to set $\Sigma_1 = \Sigma_6$ equal to the adjacency matrix of the $F_4$ diagram and use the same relations that determine all matrices $O_x$ from $O_0$, $O_1$ and $O_6$ (equivalently, 
solve the system of equations $\Sigma_x \Sigma_y = O_{xzy} \, \Sigma_z  $ with given structure constants
 $O_{xyz}$). 
 
 Another method uses our realization of $Oc(F_4)$ as  a fibered tensor product $E_{6}\otimesdot 
E_{6}$. Remember that the (left, for instance) graph of quantum symmetries is a union of four graphs $E,e,F,f$, two of type $E_6$, two of type $F_4$, so that any single connected component  describes a module action over the subalgebra associated with the first subgraph ($E$). 
 In this way we obtain four sets of matrices: $s^E_u$ (of dimension $6\times 6$), 
$s^e_u$ (of dimension $6\times 6$),  $s^F_u$ (of dimension $4 \times 4$), and $s^f_u$ (of dimension $4\times 4$). In all cases, $u$ runs from $0$ to $5$, so that these are six-elements sets.  The elements of the first set $s^E$ co\" \i ncide with the already known generators of the graph algebra of $E_6$. What we have to use in this section  is  the second module\footnote{In this respect the 
situation is similar to the analysis of $E_{7}$, which appears as a 
subgraph of its own algebra of quantum symmetries and as a module over the graph algebra of
$D_{10}$, itself a subalgebra of $Oc(E_7)$} of type $F_4$ (called $f$) and the matrices $s^f$ that 
express the
 action $E \times f \subset f$. Remember also that we have $f \times f \subset E$. 
 These matrices are as follows:
{\tiny
 $$
\begin{array}{cccccc}
s^f_0=\left(\begin{array}{cccc}1 & . & . & . \\ . & 1 & . & . \\ . & . & 1 & . \\ . & . & . & 1 \\ \end{array}\right)&
s^f_1=\left(\begin{array}{cccc}. & 1 & . & . \\ 1 & . & 2 & . \\ . & 1 & . & 1 \\ . & . & 1 & . \\ \end{array}\right)&
s^f_2=\left(\begin{array}{cccc}. & . & 2 & . \\ . & 2 & . & 2 \\ 1 & . & 2 & . \\ . & 1 & . & . \\ \end{array}\right)&
s^f_3=\left(\begin{array}{cccc}. & . & . & 2 \\ . & . & 2 & . \\ . & 1 & . & . \\ 1 & . & . & . \\ \end{array}\right)&
\end{array}
 $$
}
and $s^f_4=s^f_1$, $s^f_5=s^f_0$.
From the expressions giving the linear generators of $Oc$ as tensor products
of $E_6$ vertices, we obtain $\Sigma_{x} = s_{a}.s_{b}$, whenever $x = a \otimesdot b$.

The two methods give the same result and  we obtain the following $20$ matrices:
{\tiny
$$
\begin{array}{cccccc}
\Sigma_0 & \Sigma_1 & \Sigma_2 & \Sigma_3 & \Sigma_7 & \Sigma_8 \\
{} \\
\left(\begin{array}{cccc}
1&.&.&. \\
.&1&.&. \\
.&.&1&. \\
.&.&.&1 
\end{array}\right)
 &
\left(\begin{array}{cccc}
.&1&.&. \\
1&.&2&. \\
.&1&.&1 \\
.&.&1&. 
\end{array}\right)
&
\left(\begin{array}{cccc}
.&.&2&. \\
.&2&.&2 \\
1&.&2&. \\
.&1&.&. 
\end{array}\right)
&
\left(\begin{array}{cccc}
.&.&.&2 \\
.&.&2&. \\
.&1&.&. \\
1&.&.&. 
\end{array}\right)
&
\left(\begin{array}{cccc}
1&.&2&. \\
.&3&.&2 \\
1&.&3&. \\
.&1&.&1 
\end{array}\right)
&
\left(\begin{array}{cccc}
.&2&.&2 \\
2&.&6&. \\
.&3&.&2 \\
1&.&2&. 
\end{array}\right)
\end{array}
$$
}
and 
$ \Sigma_5 = \Sigma_{19}/2=\Sigma_{0}$,
$\Sigma_4=\Sigma_6=\Sigma_{11}= \Sigma_{15}/2=\Sigma_{18}/2=\Sigma_1$,
$\Sigma_9=\Sigma_{12}=\Sigma_{17}=\Sigma_2$,
$\Sigma_{16}=\Sigma_3$,
$\Sigma_{10}= \Sigma_{14}/2=\Sigma_7$,
$\Sigma_{13}=\Sigma_8 $. Notice that  $\Sigma_{14}, \Sigma_{15}, \Sigma_{18}, \Sigma_{19}$, as matrices over positive integers, can be divided by $2$.

As another check of the correctness of the previous calculation, there exists a relation 
that holds between matrices $F_{p}$ and matrices $\Sigma_{x}$. It could actually be used to 
determine the former from the later, although this method would be more 
complicated than the one we followed.  This relation reads:
$
F_{n} = \sum_{y} (W_{\underline{0},y})_{n,0} \, \Sigma_{y}
$
and stems from the compatibility between  
actions of $A_{11}$ and $Oc$ on the diagram $F_{4}$ and the fact 
that the unit $\underline{0}$ of $Oc$ indeed acts trivially.  

\subsection*{Sum rules}
\begin{itemize}
\item Quadratic sum rule.
Being both semi-simple and co-semi-simple, the following quadratic sum 
rule holds for the Ocneanu quantum groupoid associated with a simply 
laced diagram: $\sum_{p} d_{p}^2 = \sum_{x} d_{x}^2$, where $d_{p} = 
\sum_{a,b} (F_{p})_{a,b}$ gives the dimensions of the simple blocks 
for the first algebra structure, and $d_{x} = 
\sum_{a,b} (\Sigma_{x})_{a,b}$ gives the dimensions of the simple blocks 
for the second algebra structure (actually the algebra structure 
defined on the dual). If we take $G = E_{6}$ for instance, we get 
$d_{p}=\{6, 10, 14, 18, 20, 20, 20, 18, 14, 10, 6\}$, $d_{x}=\{6, 8, 6, 10, 14, 10, 10, 14, 10, 20, 28, 20\}$ and we check that $\sum_{p} d_{p}^2 = \sum_{x} 
d_{x}^2 = 2512$. 

In the case of the non simply laced diagram $G=F_{4}$, analogous calculations for $d_p$ and $d_x$ lead to the
following values:
\begin{eqnarray*}
d_{p}= \{4, 7, 10, 13, 14, 14, 14, 13, 10, 7, 4\} \\
d_{x}=\{4, 7, 10, 6, 7, 4, 7, 14, 20, 10, 14, 7, 10, 20, 28, 14, 6, 10, 14, 8\}
\end{eqnarray*}
Notice that  $\sum_{p} d_{p}^2 = 1256$, which is half the $E_6$ result;  this could be expected since
$F_4$ is a $\ZZ_2$ orbifold of $E_6$.  However, this value is not equal to $\sum_x d_{x}^2$. This could also be expected if
we remember that vertices $14, 15, 18, 19$ play a special role (like the ``long roots'' in the
theory of Lie algebras): these are the values for which $W_{0,x}=2 \, W_{x,0}$ (no factor $2$ for the others) and for which the $\Sigma_x$ matrices can be divided by $2$.
For this reason, we introduce another set of matrices, setting $\widetilde \Sigma_{x} = \Sigma_{x} /2$ for those four values, and equality otherwise.
We also introduce the corresponding dimensions $\widetilde d_x$, which are equal to $d_x$ except for the four special vertices where the values are divided by $2$. 
We find  $\sum_{x} d_x  \widetilde d_x = 2512$ and notice that  this value is twice the value of the sum  $\sum_{p} d_{p}^2 $.   It would be natural to introduce a quadratic form in the vector space $Oc$,  diagonal and taking the value $1$ on the basis generators $O_x$, except in positions $15, 16, 19, 20$ where the coefficients  would be equal to $2$ .  The conclusion  is that  the usual quadratic sum rule almost works, in the sense that it is  somehow twisted  by the appearance of factors $2$ which should be understood from the fact that basis generators $O_x$  of quantum symmetries have  two different lengths (corresponding to short and long roots in the theory of Lie algebras). These results are not compatible with the existence of a quantum groupo\"{\i}d structure (in the usual sense) since the  candidates for  algebras of characters associated with the two multiplicative structure --  that  would be respectively described by the semi-simple algebras $Oc$  ($20$ blocks) or the direct sum of two copies of the graph algebra  $A_{11}$( twice $11$ blocks) -- do not have the same dimension. A more general type of algebraic structure seems to be needed.

\item Linear  sum rule.
It is an observational fact (not yet understood) that the following linear sum rule also 
holds\footnote{This was noticed in \cite{PetZub:Oc}}:  $\sum_{p} d_{p} = \sum_{x} 
d_{x}$, for most $ADE$ cases; and when it does not, one also  knows how to ``correct'' the rule by introducing natural prefactors.
In the case of $E_6$, for instance, this sum equals $156$.
For the graph $F_4$ however,  $\sum_p d_p = 110$ whereas $\sum_x d_x = 220$.
This is also compatible with the previous discussion.

\item Quantum  sum rule.  
For $ADE$ diagrams $G$ with $n$ vertices $\sigma_i$ the quantum mass $m(G)$ is defined by:
$$
m(G)=\sum_{a=0}^{n-1}\left(qdim(\sigma_i)\right)^2
$$ 
where the quantum dimensions $qdim$ of the vertex $\sigma_i$ 
is given by the $i$-component of the normalized Perron-Frobenius vector, associated with the highest eigenvalue(here $\beta=\frac{1+\sqrt{3}}{\sqrt{2}}$). To get these quantities for the vertices of $Oc$, we assign $\beta$ to both chiral generators, impose that $qdim$ is an algebra morphism and use recurrence formulae for $O_x$.  

For $ADE$ cases, the following property
can be verified (see  \cite{Gil:thesis}) :  if we denote the fusion algebra of the graph $G$  by $A(G)$  (a graph algebra of type $A_n$) and the algebra of quantum symmetries by $Oc(G)$, one finds that 
 $m(A(G)) = m(Oc(G)).$  Moreover, for a graph with self-fusion, and if it is so that  $Oc(G)$ is isomorphic, as an algebra, with $G\otimes_J G$, then $m(Oc(G))=({m(G)\times m(G)}) /{m(J)}$. For instance in the $E_6$ case 
{\footnotesize
$$
m(E_6)=4(3+\sqrt{3})\quad\mbox{and}\quad m(Oc(E_6))=\frac{m(E_6)\times m(E_6)}{m(J)}=24(2+\sqrt{3}) = m(A_{11})
$$
}

However, for the non simply laced diagram $F_4$, the $qdim$ are as follows ($x=0,\dots 19$)
{\footnotesize
$$
\begin{array}{ll}
qdim\{E,e,F,f\}=&\left\{\left(1,\frac{1+\sqrt{3}}{\sqrt{2}},1+\sqrt{3},\sqrt{2},\frac{1+\sqrt{3}}{\sqrt{2}},1\right), \left(\frac{1+\sqrt{3}}{\sqrt{2}},2+\sqrt{3},\sqrt{2}(2+\sqrt{3}),1+\sqrt{3},2+\sqrt{3},\frac{1+\sqrt{3}}{\sqrt{2}}\right),\right.\\
&\\
&\left.\left(1+\sqrt{3},\sqrt{2}(2+\sqrt{3}),2(2+\sqrt{3}),\sqrt{2}(1+\sqrt{3})\right),\left(\sqrt{2},1+\sqrt{3},\sqrt{2}(1+\sqrt{3}),2\right)\right\}
\end{array}
$$
}
Like for the quadratic sum rule we introduce quantum dimensions $\widetilde{qdim}(x)$  equal to $qdim(x)$ 
except for the four vertices $14, 15, 18, 19$ where the values are divided by 2. One finds\footnote{Using the graph algebra of $F_4$ defined in the following section, one finds rather $m(F_4)=m(E_6)/2$}:
{\footnotesize
$$
 m(Oc(F_4))=\sum_{a=0}^{n-1}\left(qdim(x)\widetilde{qdim}(x)\right)^2=m(E)+m(e)+m(F)+m(f)=48(2+\sqrt{3}) = 2 m(A_{11})
$$ 
}
with $m(E)=m(f)=4(3+\sqrt{3})$ and $m(e)=m(F)=4(9+5\sqrt{3})$.
 
\item Quadratic modular double sum rule. The modular splitting relation implies the following.
Call $d^N_p = \sum_{q,r} (N_p)_{q,r}$, 
$d^{W^\prime}_x = \sum_{y,z} (W_{x,0})_{y,z}$ and
$d^{W^{\prime \prime}}_x = \sum_{y,z} (W_{0,x})_{y,z}$, then (take traces) :
$$ \sum_{p,q} d^N_p d^N_q M_{p,q} =  \sum_{x} d^{W^\prime}_x d^{W^{\prime\prime}}_x$$
Once the $W_x$ are determined, one should verify that this sum rule holds. In the simply laced case  $E_6$, 
for instance, one easily checks this identity, with
$d^{W^\prime}_x = d^{W^{\prime\prime}}_x$ for all $x$ given by  $d^W = \{20, 28, 20, 20, 28, 20, 12, 16, 12, 34, 48, 34\} $ (sum of squares is $8328$)
and $d^N =\{11, 20, 27, 32, 35, 36, 35, 32, 27, 20, 11\} $ for $A_{11}$ so that the $M$-norm square defined
by the $E_6$ modular
matrix $M$ is also  $8328$.
In the case of $F_4$  we have the same dimension vector $d^N$ with its $M$-norm square (equal to $4232$)  now defined by the $M$ matrix of $F_4$, but $d^{W\prime}_x \neq d^{W\prime\prime}$ and  we have to take 
$d^{W\prime}= \{8, 12, 16, 8, 12, 8, 12, 18, 24, 12, 18, 12, 16, 24, 16, 8, 8, 12, 8, 4\}$  together with
$ d^{W\prime\prime}= \{8, 12, 16, 8, 12, 8, 12, 18, 24, 12, 18, 12, 16, 24, 32, 16, 8, 12, 16, 8\}$ -- notice the factor 2 for entries $15, 16, 19, 20$ -- so that the right hand side of this sum rule is also $4232$, as it should.

\end{itemize}

\section{Miscellaneous comments}

\subsection*{Comparison with a direct method using self-fusion on $F_4$}

Remember that the diagram $F_{4}$ emerged from our analysis of the modular splitting equation and that we started from a given partition function. Now, we would like to reverse the 
machine and start from the diagram $F_4$ itself. We imitate
techniques initiated in \cite{Coque:Qtetra} and developed in \cite{CoqueGil:ADE}, \cite{Gil:thesis}.
\begin{itemize}
\item From the diagram, we find its adjacency matrix and call 
it $G_{1}$. 
From its eigenvalues $2 cos(r \, \pi/12)$, we find the exponents $r=1,5,7,11$. In 
particular the  highest eigenvalue is $\beta = 
\frac{1 + \sqrt{3}}{\sqrt{2}}=2 \cos(\frac{\pi}{\kappa})$ with $\kappa = 12$  gives the value
of the Coxeter number --  not the dual Coxeter number, which is different 
(and equals $9$) since the diagram is not simply laced. The quantum 
dimensions of the vertices are given by the normalized Perron-Frobenius 
vector associated with $\beta$, and we obtain the $q$-numbers $[1], 
[2], [2], [1]$, for $q = e^{\frac{i \pi}{\kappa}}$, so $q^{2\times 12}=1$.
\item The fused adjacency matrices $F_{p}$ are obtained by checking 
that the vector space of the diagram $F_{4}$ is indeed an $A_{11}$ 
module and by imposing the usual $SU(2)$ recurrence relation  for the $F_p$'s, together 
with the seed $F_0 = \one_4$ and $F_{1}=G_{1}$. 
Vertices of the  diagram $F_4$ are labelled as follows:

$$
\begin{picture}(600,80)
\put(57,65){$a_0$}
\put(87,65){$a_1$}
\put(117,65){$a_2$}
\put(147,65){$a_3$}
\multiput(60,55)(30,0){4}{\circle*{4}}
\put(60,55){\line(1,0){30}} 
\put(90,55){\vector(1,0){20}}\put(90,55){\line(1,0){30}} 
\bezier{900}(90,55)(105,75)(120,55)\put(104,65){\vector(-1,0){1}} 
\bezier{900}(90,55)(105,67)(120,55)\put(104,61){\vector(-1,0){1}} 
\put(121,55){\line(1,0){30}}
\put(57,20){$a.)\;Oriented\;graph\;F_4.$}
\put(200,55){$\sim$}
\put(257,65){$a_0$}
\put(287,65){$a_1$}
\put(317,65){$a_2$}
\put(347,65){$a_3$}

\multiput(260,55)(30,0){2}{\circle*{4}}
\multiput(320,55)(30,0){2}{\circle{4}}
\put(260,55){\line(1,0){30}}
\put(290,55){\line(1,0){28}}
\put(322,55){\line(1,0){26}}
\put(257,20){$b.)\;Coxeter-Dynkin\;F_4\;diagram.$}
\normalsize
\end{picture}
$$

\item One is tempted to analyze the possibility of defining a graph algebra structure for the diagram $F_4$.
This is indeed possible. The multiplication table given below is determined by imposing associativity, once the multiplications by $0$ (unity) and $1$ (adjacency matrix) have been defined.
$$
\begin{array}{||c||c|c|c|c||}\hline
 \cdot   &a_0&a_1&a_2&a_3 \\ 
\hline
\hline
a_0&a_0&a_1&a_2&a_3 \\ \hline
a_1&a_1&a_0+a_2&2a_1+a_3&a_2\\ \hline
a_2&a_2&2a_1+a_3&2a_2+2a_0&2a_1\\ \hline
a_3&a_3&a_2&2a_1&2a_0\\ \hline
\end{array}
$$
The graph matrices $G_a$, $a=0,1,2,3$, obtained from this table, coincide with the first four matrices $\Sigma_x$, but the reader will notice immediately that this table cannot be obtained, by restriction,  from the multiplication table of $Oc$.  Indeed, there are only  two candidates ($F$ or $f$). The second one is ruled out by the fact that $f \times f \subset E$. The first one is not a subalgebra either since $F \times F \subset  E + F$, and even if we artificially project the result (right hand side) to $F$, the obtained table will differ from the one just given. 
 The conclusion is that there is no hope to use the above graph algebra structure on $F_4$ to recover the modular matrix $M$ that we used as the starting point of the whole analysis carried out in this paper.
 Let us nevertheless proceed. 
 
 \item
 Potential candidates for the ambichiral vertices can be obtained
by imposing that the eigenvalues of the $T$ modular operator (they are well defined for the vertices of $A_{11}$) are also well defined under the induction rules (see \cite{CoqueGil:Tmodular} for details and examples). This constraint selects the set $\{a_0, a_3\}$ so that a natural guess for the corresponding algebra of quantum symmetries would be $F_4 \otimesdot F_4$ where the algebra structure of each factor  was described in the previous paragraph and where the tensor product is taken above the subalgebra generated by $\{a_0, a_3\}$. The new modular matrix $M^{new} =(W_{0,0})^{new}$ is then given by $E_{0}^{red} (E_{0}^{red} )^T$ where  the reduced essential matrix $E_{0}^{red}$ is obtained from $E_0$ by keeping only the first and last column and replacing the two others by zeros. It is equal to
{\tiny
$$
M^{new} =\left(\begin{array}{ccccccccccc}
   1 & . & . & . & 1 & . & 1 & . & . & . & 1 \\
   . & . & . & . & . & . & . & . & . & . & . \\ 
   . & . & . & . & . & . & . & . & . & . & . \\ 
   . & . & . & 1 & . & . & . & 1 & . & . & . \\ 
   1 & . & . & . & 1 & . & 1 & . & . & . & 1 \\ 
   . & . & . & . & . & . & . & . & . & . & . \\ 
   1 & . & . & . & 1 & . & 1 & . & . & . & 1 \\ 
   . & . & . & 1 & . & . & . & 1 & . & . & . \\ 
   . & . & . & . & . & . & . & . & . & . & . \\ 
   . & . & . & . & . & . & . & . & . & . & . \\ 
   1 & . & . & . & 1 & . & 1 & . & . & . & 1 \\ 
\end{array}\right)
$$
}
As expected, it differs from the matrix $M =  W_{\underline 0} =
W_{0,0}$. However it is interesting to notice that both are related by
by a conjugacy :  $
M^{new}/2= {S}^{-1}\,M\,{S}$, where
${S}$ stands for one of the two  generators of the modular group in this representation. We  find that the 
bilinear form  obtained from
$W_{00}$ is invariant under the action
of the congruence subgroup 
$S^{-1}\Gamma_0^{(2)}S=gen\left\{S^{-1}T^2S\;,\;S^{-1}(ST^{-2}S)S\right\}$
conjugated with $\Gamma_0^{(2)}$.
This can be directly verified by calculating the commutators
{\footnotesize
$$
\begin{array}{ccc}
\left[{S}^{-1} {T}_{11}^2{S}\;,\;W_{00}\right]=0&,
&
\left[{S}^{-1} {(ST^{-2}S)}_{11}{S}\;,\;W_{00}\right]=0
\end{array}
$$
}
Notice  that $M^{new}$ (which has $20$ non-zero entries) is equal to the sum of the three matrices $W_{\underline 0}$, $W_{\underline 4}$ and $W_{\underline {33^\prime}}$
associated with three self-dual points of the graph $Oc(F_4)$.
\end{itemize}
So,  we can indeed define self-fusion on the diagram $F_4$ but this associative algebra structure does not seem to be simply related with the so-called $F_4$ modular matrix.
Still another possibility would be to work with a symmetrized form of the $F_4$ diagram, \ie with an ``adjacency matrix'' that incorporates non-integer matrix elements ($q$-numbers equal to $\sqrt 2$). 
This possibility is actually quite interesting but will not be discussed here. It does not seem to allow one to recover the $F_4$ modular matrix $M$, either.

\subsection*{A relative equation of modular splitting }

 From the fact that the diagram $F_{4}$ is, geometrically, a 
$Z_{2}$ orbifold of $E_{6}$ (identify pair of vertices $(0,4)$, 
$(1,5)$ of the later), we are tempted to consider an action of the
graph algebra of $E_{6}$ (this graph has indeed self-fusion) on the 
vector space of $F_{4}$. Call ${G_{u}^{E_{6}}}$ the six $6 \times 6$ 
generators of the $E_{6}$ graph algebra and ${F_{u}^{E_{6}}}$ the six $4 
\times 4$ matrices implementing the action of $E_{6}$ on $F_{4}$.
The multiplication table of $E_6$ was given before. Its graph matrices obey the usual relations:

{\small
$
\begin{array}{ccc}
G^{E_6}_0  = \one_6 , & 
G^{E_6}_1 ,  &
G^{E_6}_2  = G^{E_6}_1 .G^{E_6}_1  - G^{E_6}_0  , \\
G^{E_6}_4  =G^{E_6}_1 .G^{E_6}_1 .G^{E_6}_1 .G^{E_6}_1 - 4 G^{E_6}_1 .G^{E_6}_1 + 2 G^{E_6}_0 ,  &
G^{E_6}_5  = G^{E_6}_1 .G^{E_6}_4  ,  &
G^{E_6}_3  =  - G^{E_6}_1 .(G^{E_6}_4  - G^{E_6}_1 .G^{E_6}_1  + 2 G^{E_6}_0) 
\end{array}
$
}

To obtain the fused matrices ${F_{u}^{E_{6}}}$ relative to this action, we set  ${F_{0}^{E_{6}}}=\one_{4}$,  ${F_{1}^{E_{6}}}= G_{1}$ (the adjacency matrix of $F_4$) and impose that 
the  ${F_{u}^{E_{6}}}$ should obey the same algebra relations as the 
${G_{u}^{E_{6}}}$.

Exactly as we had an action of $A_{11}$ on $F_{4}$, implemented by
matrices $F_{p}$, we have a 
(relative)  action of $E_{6}$ on $F_{4}$, implemented by matrices 
 ${F_{u}^{E_{6}}}$. 
 For this reason we are led to consider a ``relative'' theory of modular splitting (and a corresponding equation) with  $A_{11}$ replaced by $E_6$. In particular the graph matrices $N_{p}$ of $A_{11}$ -- the 
usual fusion matrices -- are replaced by the generators $G_u^{E_{6}}$ of the graph algebra  of $E_{6}$.
With $u,v \in E_6$. we define ``relative'' toric matrices by 
$$
u \, x \, v = \sum_y   {(W^{E_6}_{x,y})}_{u,v} \; y
$$
 The relative equation of modular splitting reads ($\tau$ is a tensorial flip):
$$\sum_{u,v} G_u^{E_{6}} \otimes G_v^{E_{6}}\; M^{rel}_{u,v} = \sum_{x \in Oc}  \tau \circ  (W^{E_6}_{x,0} \otimes 
W^{E_6}_{0,x})$$
 $M^{rel}$ describes the same $F_4$ partition function as before, but in terms of generalized characters\footnote{In our formalism, they are obtained from essential matrices of the $E_6$ diagram as described,  for instance, in \cite{CoqueMarina:minimal}}:
$$
\begin{array}{cccccc}
 \chi_{0}+\chi_{6} , &
 \chi_{1}+\chi_{5}+\chi_{7}, &
\chi_{2}+\chi_{4}+\chi_{6}+\chi_{8}, &
\chi_{3}+\chi_{5}+\chi_{9},&
\chi_{4}+\chi_{10},&
\chi_{3}+\chi_{7} \\
\end{array}
$$
If $P$ denotes the matrix of this linear transformation (it is the first essential matrix, \ie the ``intertwiner'' of the $E_6$ theory),  we have $M = P M^{rel}  P^T$.
The $E_6$ invariant, in terms of these generalized characters, with the above ordering, is diagonal and reads
 $diag(1,0,0,0,1,1)$ whereas the $F_4$ modular matrix is $M^{rel} = diag(1,0,0,0,1,0)$.
The equation of modular splitting is then solved exactly as we did in a previous section,  with the technical advantage that the size of the matrices that we have to manipulate is much smaller ($36\times 36$ rather than $121 \times 121$). Same comment for most  objects of the theory:  the analogue of $K_0$ is $36\times 20$ (rather than $121 \times 20$) and  the relative toric matrices are $6\times 6$ (rather than $11\times 11$). The twenty
generators $O_x$ are the same (their size is $20 \times 20$) and the graph of quantum symmetries is determined as before.
It is easy to translate the relative $E_6$ theory to the ``usual'' one (that uses the action of the $A_{11}$ graph algebra) by using the -- rectangular --  essential matrix $P$. It is technically easier to work with this relative theory, but the drawback is that the $E_6$ case should be analyzed first. This is why we did not follow this method in our presentation.
{\small
\paragraph{Acknowledgments} 
This work was  certainly influenced by conversations with A. Ocneanu, 
O. Ogievetsky and G. Schieber. We want to thank them here.
}

{}

\end{document}